\documentclass[aip,cha,amsmath,amssymb,preprint]{revtex4-2}

\usepackage{graphicx}
\usepackage{bm}
\usepackage{color}
\usepackage{hyperref}

\newtheorem{theorem}{Theorem}
\newtheorem{lemma}{Lemma}

\begin{document}

\title{Global Analytical Solution of the Identical Kuramoto Model for $N=3$ via Koopman Eigenfunctions}

\author{Keisuke Taga}
\email{tagakeisuke@rs.tus.ac.jp}
\affiliation{Department of Physics and Astronomy, Tokyo University of Science, Yamazaki 2641, Noda, Chiba 278-8510, Japan}

\begin{abstract}
The Kuramoto model is a paradigmatic model of collective synchronization in coupled oscillator systems.
Although its mathematical properties have been extensively investigated, explicit phase trajectories from arbitrary initial conditions have been available only for the simplest case, $N=2$.
In this study, we provide a global analytical solution for the phase trajectories of the all-to-all coupled Kuramoto model with three identical oscillators ($N=3$).
We construct Koopman eigenfunctions from products of observables whose growth rates, weighted by the exponents, sum to a constant, and obtain a nontrivial constant of motion from the ratio of two such eigenfunctions.
Using their explicit exponential time dependence, we reduce trajectory reconstruction to time-dependent quartic equations and recover the phase trajectories.
The same construction also yields Koopman eigenfunctions for the Kuramoto--Sakaguchi model with arbitrary phase lag and for the $N=4$ all-to-all and ring-coupled systems.
These results provide a concrete example in which Koopman eigenfunctions are used not only to characterize but also to explicitly solve a widely studied nonlinear model.
\end{abstract}

\maketitle

\begin{quotation}
Explicit trajectories of dynamical systems provide information that is not contained in stability, bifurcation, or asymptotic analyses alone. For the identical Kuramoto model, such trajectories are straightforward for $N=2$ but have remained difficult to obtain for $N\geq3$. Here, we obtain a global analytical solution for the smallest nontrivial case, $N=3$.

Koopman analysis describes nonlinear dynamical systems through the linear time evolution of observables. However, this linear formulation does not automatically lead to explicit solutions, since constructing suitable Koopman eigenfunctions and reconstructing the original state variables from them can both be challenging. In this work, we construct Koopman eigenfunctions explicitly and use their simple time dependence to reconstruct the individual oscillator phases algebraically. This demonstrates that Koopman eigenfunctions can be used not only to characterize nonlinear dynamics but also to obtain explicit solutions.
\end{quotation}

\section{Introduction}
\label{sec:introduction}
Synchronization is a ubiquitous collective phenomenon observed in a wide variety of natural and engineered systems~\cite{strogatz1993coupled,Acebron2005paradigm}, and it has been successfully captured by simple mathematical models.

One of the most fundamental models of synchronization is the Kuramoto model~\cite{Kuramoto1975self,Kuramoto1984chemical,Acebron2005paradigm}.
\begin{align}
\label{eq:kuramoto_model}
    \dot{\theta}_j = \omega_j + \frac{K}{N}\sum_{k=1}^N \sin(\theta_k-\theta_j),
    \qquad j=1,2,\ldots,N,
\end{align}
where $\theta_j$ is the phase of the $j$th oscillator, $\omega_j$ is its natural frequency, $K$ is the coupling strength, and $N$ is the number of oscillators.

Both in the continuum limit and in finite-dimensional settings, this model exhibits several mathematically tractable features.
For example, in the continuum limit $N\to\infty$, the onset of synchronization and the bifurcation properties of the order parameter have been analyzed for several frequency distributions~\cite{Kuramoto1984chemical,Strogatz1991Incoherence,Pazo2005uniform,Chiba2015proof}.
A further major development is the Ott--Antonsen ansatz~\cite{Ott2008LowDimensional}, which gives an invariant low-dimensional manifold and, for the Lorentzian (Cauchy) distribution, reduces the dynamics to a closed two-dimensional system.

For small-$N$ Kuramoto systems, several studies have clarified the stability, symmetry, bifurcation, and asymptotic states by analytical and numerical methods~\cite{Jadbabaie2004stabilityKuramoto,maistrenko2004desynchronization,maistrenko2005chaotic,Taylor2012noNonzeroStable,engelbrecht2014Classification,mehta2015algebraic,Ashwin2016Generalized,Xin2016analytical,Mihara2019exactKS,muller2021algebraic}.
However, obtaining explicit phase trajectories from arbitrary initial conditions remains difficult beyond the simplest case $N=2$, where the dynamics reduces to a separable differential equation.

For identical-frequency systems, Watanabe and Strogatz showed that the dynamics can be reduced to a three-dimensional system through the Watanabe--Strogatz transformation~\cite{Watanabe1993integrability,Watanabe1994constants, Pikovsky2008partially,Marvel2009identical}.
This reduction is associated with the existence of $N-3$ independent constants of motion (Watanabe--Strogatz invariants), which constrain the dynamics to three-dimensional invariant manifolds.

{For Kuramoto--Sakaguchi systems, the symmetry under a uniform phase shift further reduces the essential dynamics to two dimensions~\cite{Chen_2017}. In the special case of pure cosine coupling, an additional first integral attributed to Swift~\cite{Watanabe1993integrability} exists which can be written as
\begin{align}
I_{\mathrm S} = \prod_{j=1}^{N}
\sin\frac{\theta_j-\theta_{j+1}}{2},
\qquad
\theta_{N+1}=\theta_1 .
\end{align}
Together with the two-dimensional reduction, this additional conserved quantity makes the reduced dynamics completely integrable~\cite{Watanabe1993integrability,Watanabe1994constants}.

For the standard sine-coupled Kuramoto model, however, the Watanabe--Strogatz invariants and the reduction associated with the uniform phase-shift symmetry do not by themselves imply complete integrability or provide an explicit reconstruction of the individual phase trajectories. Indeed, after removing the uniform phase rotation, the reduced dynamics is described by a two-dimensional gradient flow~\cite{Chen_2017}, in contrast to the completely integrable structure of the pure cosine-coupling case~\cite{Watanabe1993integrability,Watanabe1994constants,Ashwin2016Generalized}.
}

In the present work, we construct Koopman eigenfunctions and use them to obtain the individual phase trajectories as explicit functions of time.
In recent decades, the Koopman framework~\cite{Koopman1931hamiltonian,Koopman1932spectra,mezic2005spectral,budivsic2012applied,brunton2022modern} has attracted increasing attention in the analysis of nonlinear dynamical systems.
Instead of the state variables themselves, this framework focuses on observables, namely functions of the state variables.
The Koopman operator governs the time evolution of observables. Because this operator is linear, nonlinear dynamics can be analyzed from a linear perspective.

\begin{figure}[h]
    \centering
    \includegraphics[width=0.85\linewidth]{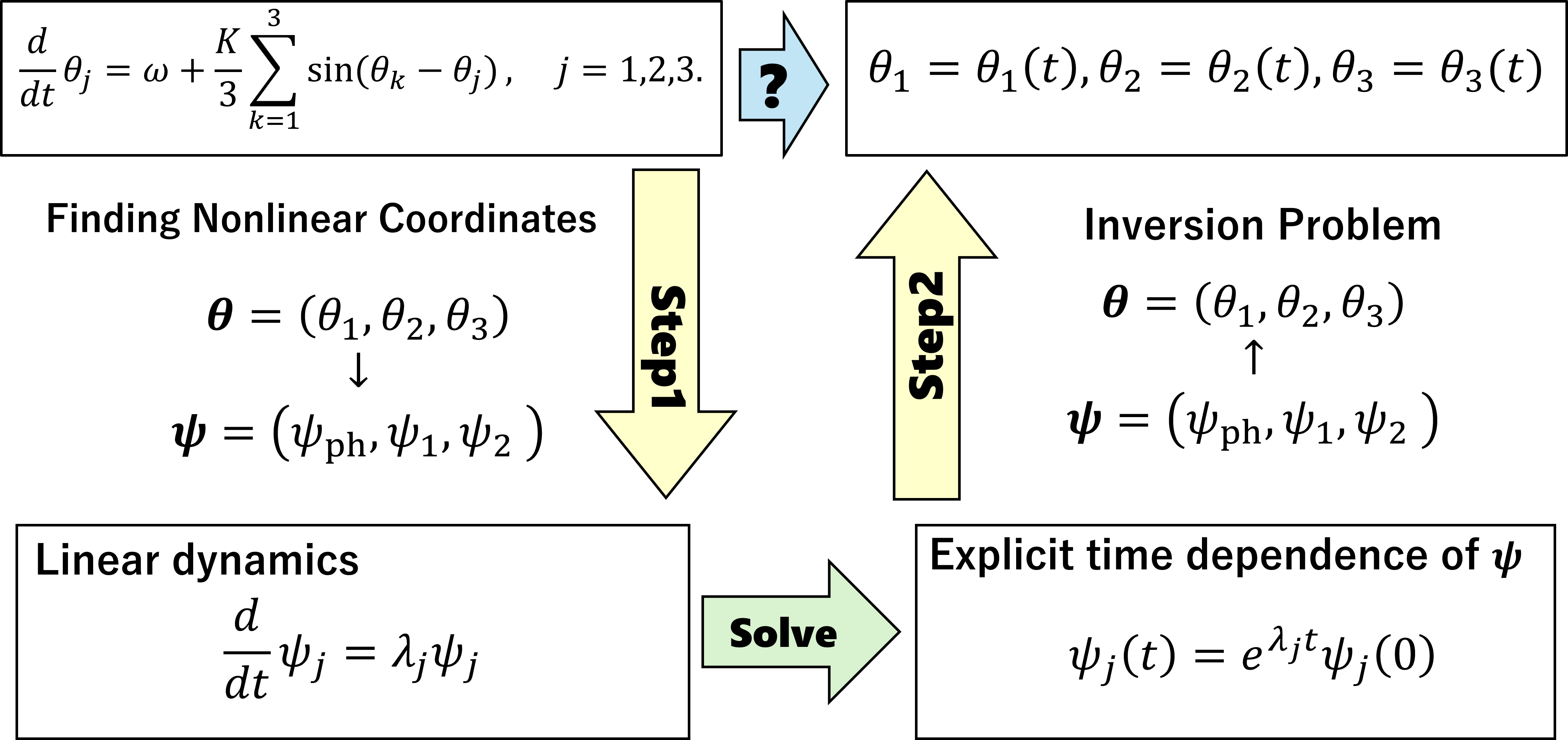}
    \caption{Procedure for obtaining the exact trajectories of the identical Kuramoto model with $N=3$.
    Step~1: the phases $\bm{\theta}$ are mapped to the Koopman eigenfunctions $\bm{\psi}=(\psi_{\mathrm{ph}},\psi_1,\psi_2)$ constructed in Sec.~\ref{sec:eigenfunctions_3KM}, in which the dynamics becomes linear with eigenvalues $\lambda_{\mathrm{ph}}=i\omega$ and $\lambda_1=\lambda_2=-K$.
    Step~2: the explicit time dependence $\psi_j(t)=e^{\lambda_j t}\psi_j(0)$ is inverted to recover $\bm{\theta}(t)$ (Sec.~\ref{sec:derivation}).}
    \label{fig:schematic}
\end{figure}

{The present work has two main components, summarized in Fig.~\ref{fig:schematic}.
First, we construct Koopman eigenfunctions that provide three independent relations between the phases and time for the Kuramoto model with identical oscillators for $N=3$ (Step~1; Sec.~\ref{sec:eigenfunctions_3KM}).

Second, we use these eigenfunctions to reconstruct the explicit phase trajectories by reducing the inverse problem to time-dependent quartic equations.
The coefficients of these quartic equations are polynomial functions of the Koopman eigenfunctions, and the algebraic branch specified by the initial condition recovers the corresponding phase trajectory (Step~2; Sec.~\ref{sec:derivation}).
As further extensions, Koopman eigenfunctions for the Kuramoto--Sakaguchi model and for systems with $N\geq 4$ are discussed in Sec.~\ref{sec:extensions}.}

\section{Koopman analysis}
\label{sec:eigenfunction}
\subsection{Koopman eigenfunctions and growth rates}
The Koopman framework was originally introduced by Koopman and von Neumann~\cite{Koopman1931hamiltonian,Koopman1932spectra} in an operator-theoretic formulation of classical dynamics. It has since been developed as a useful viewpoint for nonlinear dynamical systems, including dissipative systems~\cite{mezic2005spectral,budivsic2012applied,lasota2013chaos,mezic2013analysis,mauroy2013isostables,Wilson2016isostable,Mauroy2016global,shirasaka2017phase,mauroy2020koopman,nakao2020spectral,Wilson2020paradigm,Kaiser2021KoopmanControl,taga2021koopman,brunton2022modern,taga2024dynamic,Morrison2024koopman,yawata2024autoencoder,yawata2026Global}, and has also been applied to the Kuramoto model~\cite{susuki2016power,hu2020synchronization,wang2021probing,mihara2022basin,thibeault2025kuramoto,Taga2026Liouvillian}. In the present work, we focus on Koopman eigenfunctions, since they allow us to relate the state variables directly to time.

Koopman analysis focuses on the time evolution of observables, namely scalar functions $u=u(\bm{x})$, rather than on the state variable $\bm{x}$ itself.
{In the present analysis, we consider complex-valued observables $u:\mathcal D\to\mathbb C$ that are continuously differentiable on an open region $\mathcal D$ where the observable is regular.}
The \textit{Koopman generator} $\mathcal K$, that is, the infinitesimal form of the Koopman operator, is defined by 
\begin{align}
    \frac{d u}{dt}
    = \bm{F}(\bm{x})\cdot\nabla u
    =: \mathcal{K}u,\quad \dot{\bm{x}} = \bm{F}(\bm{x}).
\end{align}
Although the vector field $\bm{F}$ is generally nonlinear, the Koopman generator is linear, so nonlinear dynamics can be analyzed from a linear perspective.

An observable $\psi_\lambda$ and a scalar $\lambda\in\mathbb{C}$ satisfying
\begin{align}
    \label{eq:koopman_eigenfunction}
    \frac{d \psi_\lambda}{dt}
    = \mathcal{K}\psi_\lambda = \lambda \psi_\lambda
\end{align}
are called a \textit{Koopman eigenfunction} and the corresponding \textit{Koopman eigenvalue}, respectively.
For a general observable $u$ that is not an eigenfunction, the quantity
\begin{align}
    \Lambda(\bm{x}):= \frac{\mathcal{K}u(\bm{x})}{u(\bm{x})},
\end{align}
is state-dependent rather than constant.
We refer to $\Lambda$ as the \textit{pointwise growth rate} of $u$ in the following discussion.

For an observable of the form
\begin{align}
\label{eq:ob_prod}
u = \prod_{j=1}^M u_j^{\alpha_j},\quad \{\alpha_j\}_{j=1,2,\ldots,M}\in \mathbb{C}^M,
\end{align}
the corresponding growth rate $\Lambda$ is given by
\begin{align}
\label{eq:lambda_sum}
\Lambda = \sum_{j=1}^M \alpha_j \Lambda_j,
\end{align}
where $\Lambda_j$ are the growth rates associated with each observable $u_j$.
Hence, if a suitable linear combination of the growth rates is constant, the corresponding product gives a Koopman eigenfunction~\cite{Morrison2024koopman}. We use this property to construct Koopman eigenfunctions of the system.

{We note that this construction is closely related to the Darboux-theoretic construction of first integrals, in which products of Darboux polynomials are combined so that a weighted sum of their cofactors vanishes~\cite{Darboux1878Memoire,zhang2017integrability}.
The present construction may be viewed as a generalization of this idea in the sense that, instead of requiring the linear combination of the growth rates to vanish, we require it to be constant, thereby obtaining a Koopman eigenfunction~\cite{Morrison2024koopman}.
}

A Koopman eigenfunction directly provides a relation between the state variables and time,
\begin{align}
\psi_\lambda(\bm{x}(t)) = e^{\lambda t}\psi_\lambda(\bm{x}(0)).
\end{align}
If sufficiently many such relations of this form are available, the state variables can, in principle, be reconstructed exactly as functions of time by solving these relations.

\subsection{Koopman eigenfunctions for the identical Kuramoto model with $N=3$}
\label{sec:eigenfunctions_3KM}
The identical Kuramoto model for $N=3$ is given by
\begin{align}
\label{eq:3kuramoto}
    \dot{\theta}_j=\omega +\frac{K}{3}\sum_{k=1}^3\sin(\theta_k-\theta_j),\quad j=1,2,3.
\end{align}
{Throughout this paper, we take $K,\omega\in\mathbb R$ and consider the dynamics of real-valued phases $\theta_j\in\mathbb R$, which are physically interpreted modulo $2\pi$.}
For Eq.~\eqref{eq:3kuramoto}, the Koopman generator acts on an observable $u(\bm{\theta})$ as
\begin{align}
    \mathcal{K}u = \sum_{j=1}^{3}
    \left(
    \omega+\frac{K}{3}\sum_{k=1}^{3}\sin(\theta_k-\theta_j)
    \right)
    \frac{\partial u}{\partial \theta_j}.
\end{align}
To derive exact trajectories, we seek three independent relations connecting the three phases to time.

First, a trivial Koopman eigenfunction follows from the fact that the mean phase of the Kuramoto model~\eqref{eq:3kuramoto} evolves with constant velocity:
\begin{align}
\label{eq:mean_dynamics}
    \dot{\Theta} = \omega,\quad \Theta:=\frac{1}{3}\sum_{j=1}^3 \theta_j,
\end{align}
so that
\begin{align}
\label{eq:phase_rel}
   \Theta(t) = \omega t + \frac{1}{3}\sum_{j=1}^3 \theta_j(0).
\end{align}
This gives the first required relation between the phases and time.
Although we will use Eq.~\eqref{eq:phase_rel} directly in Sec.~\ref{sec:derivation}, we obtain the corresponding Koopman eigenfunction as
\begin{align}
    \psi_{\mathrm{ph}} = e^{i\Theta}, \qquad \mathcal{K}\psi_{\mathrm{ph}} = i\omega\,\psi_{\mathrm{ph}}.
\end{align}

For identical oscillator systems of the form considered below, including the identical Kuramoto model, Watanabe--Strogatz theory gives $N-3$ invariants for $N\geq4$~\cite{Watanabe1994constants,Marvel2009identical,Taga2026Liouvillian}.
These invariants are Koopman eigenfunctions with zero eigenvalue.
In the present case $N=3$, however, no Watanabe--Strogatz invariant is available.

To construct the nontrivial Koopman eigenfunctions, we apply the construction based on growth rates. 
We use the following lemma, which applies whenever the pairwise difference dynamics has a suitable trigonometric form.

\begin{lemma}
\label{lem:pointwise}
Consider variables $\varphi_j$ and $\varphi_k$ satisfying
\begin{align}
    \label{eq:identical_diff}
    \dot{\varphi}_j-\dot{\varphi}_k = g(\bm{\varphi})(\cos\varphi_j-\cos\varphi_k) + h(\bm{\varphi})(\sin\varphi_j-\sin\varphi_k),
\end{align}
where $\bm{\varphi}=(\varphi_1,\varphi_2,\ldots,\varphi_N)$.
Then the growth rate $\Lambda_{jk}$ of the observable
\begin{align}
    u_{jk}=\sin\frac{\varphi_j-\varphi_k}{2}
\end{align}
is given by
\begin{align}
\Lambda_{jk} = -\frac{g(\bm{\varphi})}{2}\bigl(\sin\varphi_j+\sin\varphi_k\bigr) + \frac{h(\bm{\varphi})}{2}\bigl(\cos\varphi_j+\cos\varphi_k\bigr).
\end{align}
\end{lemma}
A proof is given in Appendix~\ref{sec:proof_pointwise}.

{Suppose that a set of oscillators obeys equations of the form
\begin{align}
    \label{eq:identical_oscillators}
    \dot{\varphi}_j = f(\bm{\varphi})
    + g(\bm{\varphi})\cos\varphi_j
    + h(\bm{\varphi})\sin\varphi_j
\end{align}
with common functions $f$, $g$, and $h$.
Then any pair of oscillators in this set automatically satisfies Eq.~\eqref{eq:identical_diff}.
Therefore, Eq.~\eqref{eq:identical_oscillators} with common $f$, $g$, and $h$ gives a sufficient condition for Lemma~\ref{lem:pointwise}.}

For the identical Kuramoto model with $N=3$, Lemma~\ref{lem:pointwise} applies in two ways.
First, Eq.~\eqref{eq:3kuramoto} itself belongs to the class of Eq.~\eqref{eq:identical_oscillators}, yielding growth rates for observables of the form $\sin(\Delta_j/2)$.
Second, the phase-difference dynamics has the form
\begin{align}
    \label{eq:diff_KM}
    \dot{\Delta}_j = G(\Delta_1,\Delta_2,\Delta_3)-K\sin\Delta_j,
\end{align}
where $\Delta_1=\theta_1-\theta_2$, $\Delta_2=\theta_2-\theta_3$, $\Delta_3=\theta_3-\theta_1$, and
\begin{align}
G(\Delta_1,\Delta_2,\Delta_3)
=
\frac{K}{3}
(\sin\Delta_1+\sin\Delta_2+\sin\Delta_3)
\end{align}
is common to every $\Delta_j$.
Thus, Lemma~\ref{lem:pointwise} can be applied again, now yielding growth rates for observables of the form $\sin((\Delta_j-\Delta_k)/2)$.
Suitable linear combinations of the growth rates from these two families yield Koopman eigenfunctions.
{For example, let $\Lambda_1$, $\Lambda_4$, and $\Lambda_6$ denote the growth rates of $\sin(\Delta_1/2)$, $\sin((\Delta_1-\Delta_2)/2)$, and $\sin((\Delta_3-\Delta_1)/2)$, respectively. Applying Lemma~\ref{lem:pointwise}, we find
\begin{align*}
3\Lambda_1-\Lambda_4-\Lambda_6=-K.
\end{align*}
This yields the Koopman eigenfunction $\psi_1$ below. The other two eigenfunctions follow by cyclic permutation.}

\begin{theorem}
\label{thm:eigenfunctions_N3}
For the identical three-oscillator Kuramoto model, the functions
\begin{align}
\label{eq:koopman_eigenfunction_phase_original}
    \psi_1 = \frac{\sin^3\frac{\Delta_1}{2}}{\sin\frac{\Delta_1-\Delta_2}{2}\ \sin\frac{\Delta_3-\Delta_1}{2}},\quad
    \psi_2 = \frac{\sin^3\frac{\Delta_2}{2}}{\sin\frac{\Delta_2-\Delta_3}{2}\ \sin\frac{\Delta_1-\Delta_2}{2}},\quad
    \psi_3 = \frac{\sin^3\frac{\Delta_3}{2}}{\sin\frac{\Delta_3-\Delta_1}{2}\ \sin\frac{\Delta_2-\Delta_3}{2}}
\end{align}
satisfy
\begin{align}
\mathcal{K}\psi_j=-K\psi_j,
\qquad j=1,2,3,
\end{align}
and obey the linear relation
\begin{align}
\psi_1+\psi_2+\psi_3=0.
\end{align}

Together with the phase Koopman eigenfunction $\psi_{\mathrm{ph}}=e^{i\Theta}$, any two of $\psi_1,\psi_2,\psi_3$ form {a locally functionally independent set on the region where
\begin{align}
\sin\frac{\Delta_j}{2}\neq0,\qquad
\sin\frac{\Delta_j-\Delta_k}{2}\neq0
\quad (j\neq k).
\end{align}}
\end{theorem}

Further details of the derivation are given in Appendix~\ref{sec:eig_deriv}, and the linear relation is proved in Appendix~\ref{sec:proof_sum}. The local functional independence is established in Appendix~\ref{sec:invertibility}.

Together with $\psi_{\mathrm{ph}}$, these eigenfunctions define nonlinear coordinates for the identical Kuramoto model with $N=3$, in which the dynamics is reduced to the linear system
\begin{align}
\label{eq:linear_ode}
   \frac{d}{dt}
\begin{pmatrix}
 \psi_{\mathrm{ph}} \\ \psi_1 \\ \psi_2 
\end{pmatrix}
=
\begin{pmatrix}
 i\omega & 0 & 0\\
 0 & -K & 0\\
 0 & 0 & -K
\end{pmatrix}
\begin{pmatrix}
 \psi_{\mathrm{ph}} \\ \psi_1 \\ \psi_2 
\end{pmatrix}.
\end{align}
Solving Eq.~\eqref{eq:linear_ode} yields
\begin{align}
    \psi_{\mathrm{ph}}(t) = e^{i\omega t}\ \psi_{\mathrm{ph}}(0),\quad \psi_1(t) = e^{-K t}\ \psi_1(0),\quad \psi_2(t) = e^{-K t}\ \psi_2(0).
\end{align}

{We note that, since $\psi_1$ and $\psi_2$ share the same eigenvalue $-K$, their ratio
\begin{align}
    \psi_{\mathrm{inv}}:=\frac{\psi_1}{\psi_2}
\end{align}
satisfies $\mathcal{K}\psi_{\mathrm{inv}}=0$ and thus defines a nontrivial invariant.
Combining $\psi_{\mathrm{ph}}$, $\psi_{1,2}$, and $\psi_{\mathrm{inv}}$, the dynamics can be viewed in terms of phase, amplitude, and invariant components, as discussed in Sec.~\ref{sec:reduction}.}

In the next section, we use $\psi_1$ and $\psi_2$, together with the mean-phase relation for $\Theta$, to derive exact trajectories.

%%%%%%%%%%%%%%%%%%%%%%%%%%%%%%%%%%%%%
\section{Derivation of exact trajectories for the identical Kuramoto model with $N=3$}
% \subsection{Trajectory reconstruction}
\label{sec:derivation}
In this section, we use the Koopman eigenfunctions to express the phase trajectories through time-dependent quartic equations and to select the algebraic branch corresponding to the initial condition.
We begin with $\psi_1$ and $\psi_2$ in Eq.~\eqref{eq:koopman_eigenfunction_phase_original}. 
{We first consider the generic case, in which $\psi_1\psi_2\psi_3\neq0$, $\psi_j\neq\psi_k$ for $j\neq k$, and all denominators of the Koopman eigenfunctions are nonzero.
The remaining cases are summarized at the end of this section (Table~\ref{tab:cases}).}

For the algebraic analysis, we introduce new variables $X_j = e^{i\Delta_j}$. By substituting them into Eq.~\eqref{eq:koopman_eigenfunction_phase_original}, we obtain the following algebraic equations in three variables.
\begin{align}
\label{eq:koopman_polynomial}
2i\psi_1\,X_1^{\frac{1}{2}}\,(X_1-X_2)\,(X_3-X_1) - (X_1-1)^{3}\,X_2^{\frac{1}{2}}X_3^{\frac{1}{2}}
=0,\cr
2i\psi_2\,X_2^{\frac{1}{2}}\,(X_2-X_3)\,(X_1-X_2) - (X_2-1)^{3}\,X_1^{\frac{1}{2}}X_3^{\frac{1}{2}}
=0,
\end{align}
where $X_j^{\frac{1}{2}}=e^{i\frac{\Delta_j}{2}}$.
To reduce these equations to two variables, we use $X_1X_2X_3=1$, which follows from $\Delta_1+\Delta_2+\Delta_3=0\pmod{2\pi}$, and eliminate $X_3=1/(X_1X_2)$.
After clearing denominators, we obtain the following polynomials, denoted by (P1) and (P2):
\begin{align}
    2i\psi_1\,(X_1-X_2)\,(1-X_1^{2}X_2) - (X_1-1)^{3}\,X_2=0, \tag{P1}\\
    2i\psi_2\,(X_2-X_1)\,(1-X_1X_2^{2}) - (X_2-1)^{3}\,X_1=0.\tag{P2}
\end{align}

To obtain one-variable polynomial equations for $X_1$ and $X_2$, we eliminate $X_2$ and $X_1$ from (P1) and (P2), respectively, using resultants.
This gives
\begin{align}
    \mathrm{Res}_{X_2}(\mathrm{P1},\mathrm{P2})
    &= X_1^2(X_1-1)^5 R_1(X_1)=0,\tag{Q1}\\
    \mathrm{Res}_{X_1}(\mathrm{P1},\mathrm{P2})
    &= X_2^2(X_2-1)^5 R_2(X_2)=0.\tag{Q2}
\end{align}
{Since $X_j=e^{i\Delta_j}$ is nonzero, the factors $X_j^2$ do not correspond to phase differences. The factors $(X_j-1)^5$ are associated with the degenerate cases $\Delta_j=0\pmod{2\pi}$, which are treated separately in Appendix~\ref{sec:degenerate_cases}. We therefore focus on the remaining factors $R_j$, which are quartic polynomials given by}
\begin{align}
\label{eq:quartic_R}
    R_j(X_j) = a_{j,4}X_j^4+a_{j,3}X_j^3+a_{j,2}X_j^2
    +a_{j,1}X_j+a_{j,0}.
\end{align}
{The coefficients $a_{j,k}$ are polynomial functions
of the Koopman eigenfunctions and are given explicitly by
\begin{align}
\label{eq:quartic_R_coefficient}
a_{j,4}
&= -(2\psi_j-i)^2
(2\psi_{j+1}+i)(2\psi_{j+2}+i),\cr
a_{j,3}
&= -4(\psi_j+i)(2\psi_j-i)
(2\psi_{j+1}+i)(2\psi_{j+2}+i),\cr
a_{j,2}
&= -6\bigl(
8\psi_j^2\psi_{j+1}\psi_{j+2}
+2\psi_j^2+2\psi_{j+1}^2
+2\psi_{j+2}^2+1
\bigr),\cr
a_{j,1}
&= -4(\psi_j-i)(2\psi_j+i)
(2\psi_{j+1}-i)(2\psi_{j+2}-i),\cr
a_{j,0}
&= -(2\psi_j+i)^2
(2\psi_{j+1}-i)(2\psi_{j+2}-i),
\end{align}
where the indices are understood cyclically modulo $3$.}

Substituting $\psi_j(t)=e^{-Kt}\psi_j(0)$ into the coefficients,
we obtain time-dependent quartic equations for $X_1(t)$ and $X_2(t)$.
Their roots can be expressed algebraically, for example by Ferrari's method, although the resulting radical expressions are too cumbersome to display here.

{Only roots on the unit circle correspond to real phase differences.
Mapping the unit circle to the real axis by $c_j=\cot(\Delta_j/2)$, the discriminant analysis in Appendix~\ref{sec:quartic_roots} shows that $R_j$ has exactly two unit-circle roots in the generic case, and that roots can coincide only if $\psi_j=0$ or $\psi_j=\psi_k$ for some $j\neq k$.
Since all $\psi_j$ evolve with the common factor $e^{-Kt}$, these conditions never arise at finite time; hence, the pair of unit-circle roots of $R_1$ and $R_2$ selected at $t=0$ by (P1), (P2), and the initial condition can be followed uniquely for all time.
A fixed radical expression of the roots may jump to another root at a branch cut, in which case the selected root is represented by another radical expression that is continuous with the previous one.}

The phase differences are then recovered from the selected unit-circle roots as
\begin{align}
    \Delta_1(t) = -i\,\log X_1(t),\qquad
    \Delta_2(t) = -i\,\log X_2(t),
\end{align}
where the branches of the logarithm are chosen so that $\Delta_j(t)$ agrees with the given initial condition at $t=0$ and is continuous in $t$.
Thus, the individual phases $\theta_1$, $\theta_2$, and $\theta_3$ are recovered as follows.
\begin{align}
\label{eq:solution}
    \theta_1 = \Theta + \frac{2\Delta_1+\Delta_2}{3} ,\quad \theta_2 = \Theta + \frac{\Delta_2 - \Delta_1}{3},\quad \theta_3 = \Theta + \frac{-\Delta_1-2\Delta_2}{3}.
\end{align}
This gives the exact trajectories of the identical Kuramoto model for $N=3$.
 
\begin{figure}[h]
    \centering
    \includegraphics[width=0.5\linewidth]{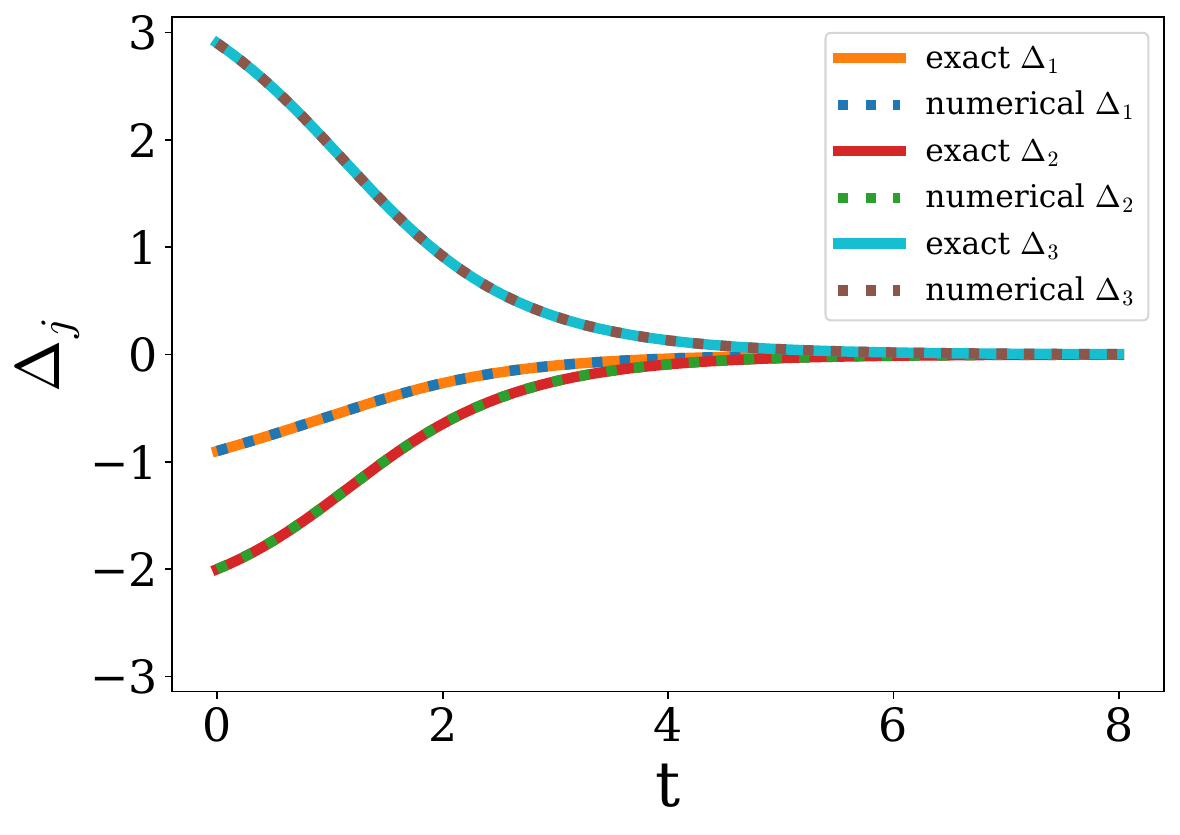}\includegraphics[width=0.5\linewidth]{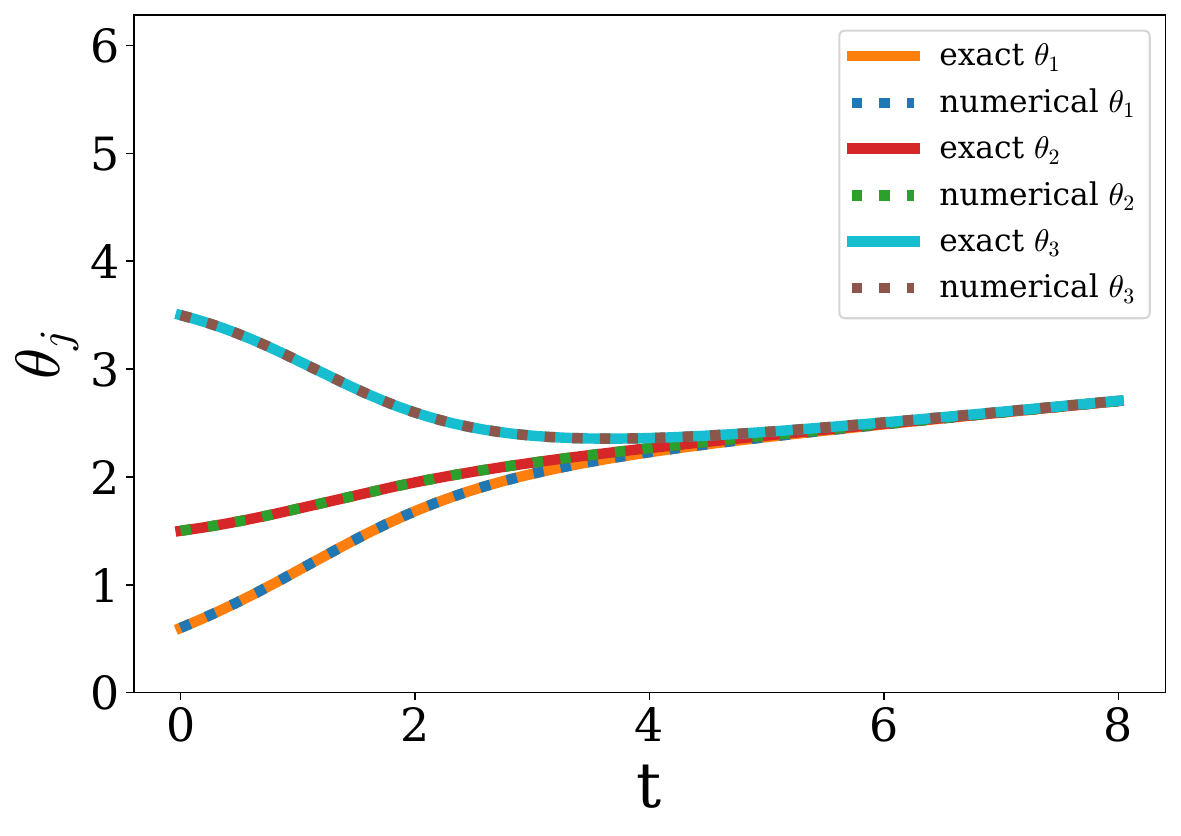}\\
    \hspace{1cm}(a)\hspace{7.5cm}(b)
    \caption{Comparison of the exact and numerical phase dynamics of the identical Kuramoto model for $N=3$ with $K=1$, $\omega =\frac{\pi}{20}$, initial phases $(\theta_1,\theta_2,\theta_3)=(0.6,\ 1.5,\ 3.5)$, and $t\in[0,\ 8]$. Solid lines represent the trajectories reconstructed from the Koopman eigenfunctions and the quartic equations {$R_j(X_j)=0$ defined by Eq.~\eqref{eq:quartic_R}}, and dotted lines show direct numerical integration as a visual reference. (a) The trajectories of the phase differences $\Delta_{1,2,3}$. (b) The trajectories of the original phases $\theta_{1,2,3}$.}
    \label{fig:N=3}
\end{figure}

Figure~\ref{fig:N=3} illustrates the reconstructed trajectories. 
Direct numerical integration of the original Kuramoto model is also shown as a visual reference.
An additional discussion on the selected and non-selected branches is given in Sec.~\ref{sec:branch}.

{The remaining initial conditions are included in exceptional sets of measure zero, which are treated in the appendices as summarized in Table~\ref{tab:cases}.
Each case in Table~\ref{tab:cases} defines an invariant set of the dynamics, so that an initial condition remains in the same case for all time.
These cases are mutually disjoint and together cover all initial conditions. Thus, the resulting trajectories provide a global analytical solution, valid for arbitrary initial conditions and for all time.
    
\begin{table}[h]
\caption{Classification of initial conditions. The cases are mutually disjoint and cover all initial conditions, and each defines an invariant set of the dynamics.}
\label{tab:cases}
\begin{ruledtabular}
\begin{tabular}{p{7cm} p{3cm}}
Condition & Details \\
\hline
\multicolumn{2}{l}{\textit{Regular: all denominators of $\psi_{1,2,3}$ are nonzero}} \\
$\psi_1\psi_2\psi_3\neq0$, $\psi_j\neq\psi_k$ (generic) & Sec.~\ref{sec:derivation} \\
$\psi_j=0$, i.e., $\Delta_j=0 \pmod {2\pi}$ & Apx.~\ref{sec:degenerate_cases}, Case 1 \\
$\psi_{j+1}=\psi_{j+2}\neq0$ & Apx.~\ref{sec:degenerate_cases}, Case 2 \\
\hline
\multicolumn{2}{l}{\textit{Singular: $\Delta_j=\Delta_{j+1}=\delta \pmod {2\pi}$ for some $j$}} \\
Phase-locked states, $\delta\in\{0,\pi,\pm2\pi/3\}$ & Apx.~\ref{sec:singular_cases}, Case 1 \\
Otherwise & Apx.~\ref{sec:singular_cases}, Case 2 \\
\end{tabular}
\end{ruledtabular}
\end{table}
    }

%%%%%%%%%%%%%%%%%%%%%%%%%%%%%%%%
\section{Discussion}
\label{sec:discussion}

\subsection{The four algebraic branches}
\label{sec:branch}
In Sec.~\ref{sec:derivation}, one of the four roots of each $R_j$ was selected by the initial condition. Here we discuss the remaining roots.

\begin{figure}[h]
    \centering
    \includegraphics[width=0.8\linewidth]{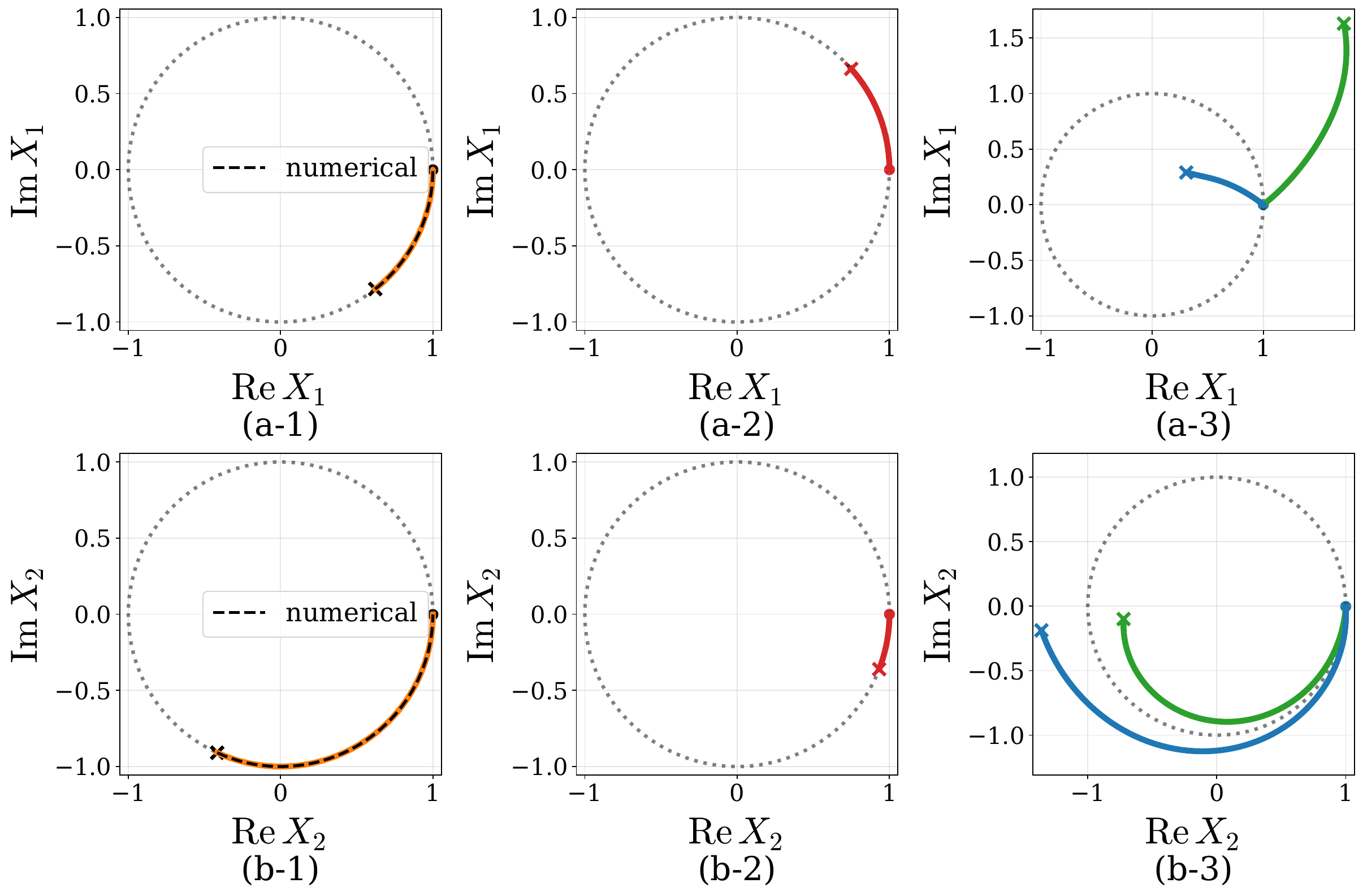}
    \caption{Candidate branches obtained from the quartic reconstruction of $X_1=e^{i\Delta_1}$ and $X_2=e^{i\Delta_2}$ for $K=1$, $(\theta_1(0),\theta_2(0),\theta_3(0))=(0.6,1.5,3.5)$, and $t\in[0,8]$. (a, b) Four candidate branches for (a) $X_1$ and (b) $X_2$, together with the numerical trajectories. Crosses and solid circles indicate the initial and final points, respectively. (a-1, b-1) Unit-circle branches selected by the initial condition. (a-2, b-2) The other unit-circle branches. (a-3, b-3) Off-circle branches forming reciprocal-conjugate pairs. Branches with the same color form paired roots satisfying (P1) and (P2) simultaneously.}
    \label{fig:N=3_unit_circle}
\end{figure}

\subsubsection{Unit-circle branches}
As shown in Sec.~\ref{sec:derivation}, two of the four roots lie on the unit circle. We first discuss these two unit-circle branches.
As indicated by the black dashed trajectories in Fig.~\ref{fig:N=3_unit_circle}(a-1) and (b-1), one branch is consistent with the given initial condition and agrees with the numerical trajectory.
Another unit-circle branch appears in Fig.~\ref{fig:N=3_unit_circle}(a-2) and (b-2).
To further characterize these two branches, we consider the energy of the corresponding classical XY model.

After moving to the uniformly rotating frame, the identical all-to-all Kuramoto model with $N=3$ is the gradient flow of a three-site classical XY model without an external field~\cite{stanley1968nVector,Acebron2005paradigm}, with Hamiltonian
\begin{align}
\label{eq:XY_hamiltonian}
    H=-J\sum_{j=1}^{3}\cos(\theta_{j+1}-\theta_j),
    \qquad \theta_4=\theta_1,
\end{align}
where $J=K/3$.

\begin{figure}[h]
\centering
    \includegraphics[width=0.9\linewidth]{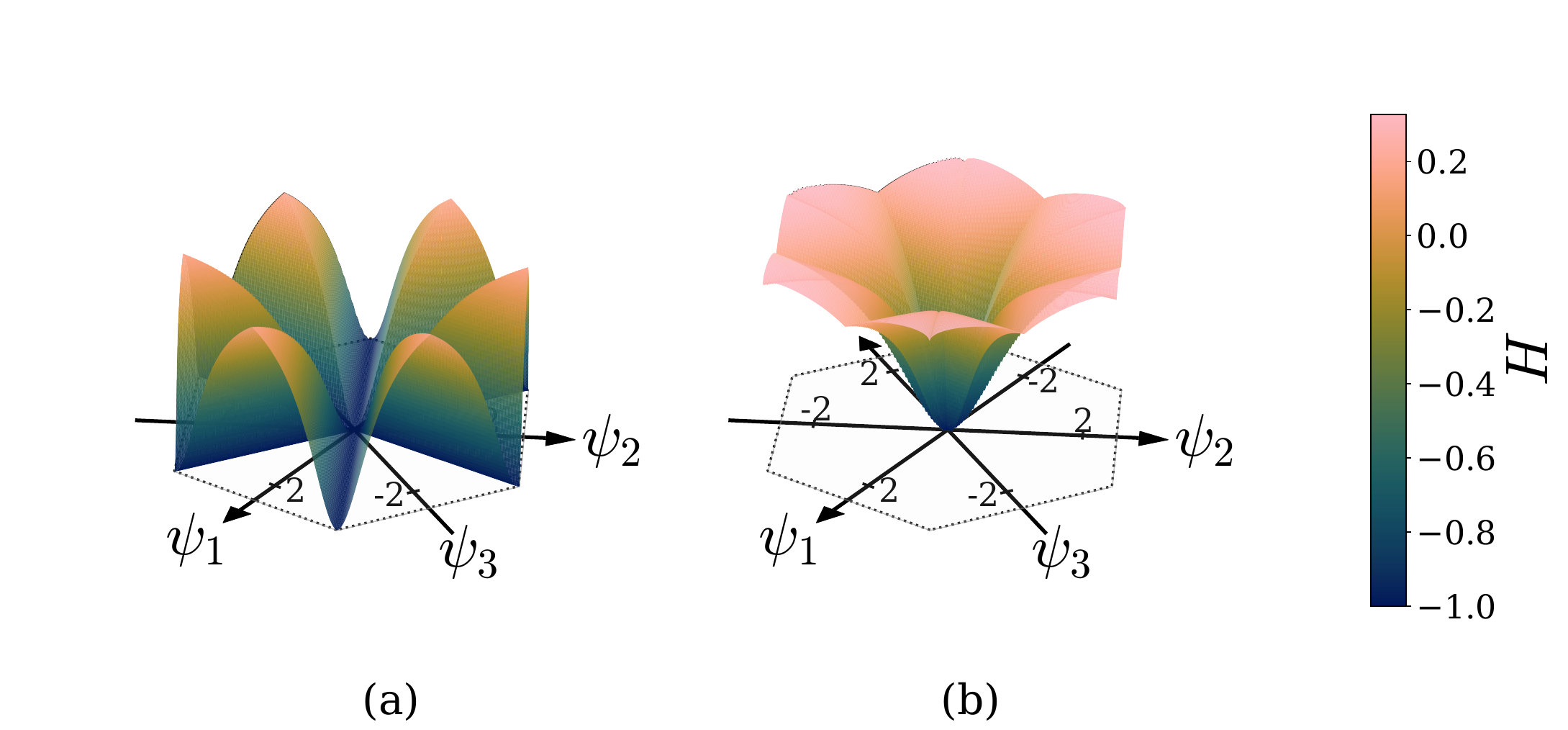}
    \caption{Two real energy sheets of the XY Hamiltonian $H$ for $3J=K=1$, represented in projected Koopman coordinates satisfying $\psi_1+\psi_2+\psi_3=0$.
    The sheets touch along $\psi_j=\psi_k$, where the quartic equation for $H$ has a multiple root.
    (a) Lower-energy sheet. (b) Higher-energy sheet.}
    \label{fig:landscape}
\end{figure}

Eliminating the phase variables yields a quartic equation for $H$ whose coefficients are polynomials in $\psi_1$ and $\psi_2$ (see Appendix~\ref{sec:xy_quartic}).
Generically, this equation has two real roots, corresponding to the lower- and higher-energy sheets shown in Fig.~\ref{fig:landscape}.
{The other unit-circle branch is itself a trajectory of the Kuramoto model, starting from the other phase configuration that shares the same values of $\psi_1$ and $\psi_2$; the two configurations are distinguished by their values of $H$.}
A complete characterization of which regions of the phase-difference domain correspond to the lower- and higher-energy sheets remains open.

\subsubsection{Off-circle branches}
The remaining roots in Fig.~\ref{fig:N=3_unit_circle}(a-3) and (b-3) form reciprocal-conjugate pairs.
Specifically, if $X_j$ is a root of Eq.~\eqref{eq:quartic_R}, then $\bar{X}_j^{-1}$ is also a root, where overbar denotes the complex conjugate. Therefore the paired branches are related by $X_j\mapsto\bar{X}_j^{-1}$.

The corresponding complexified dynamics is naturally expressed in terms of $X_j=e^{i\Delta_j}$.
From Eq.~\eqref{eq:diff_KM},
\begin{align}
\label{eq:riccati}
    \dot{X}_j
    = \frac{K}{6}\sum_{k=1}^3
      \left(X_k-\frac{1}{X_k}\right)X_j
      -\frac{K}{2}(X_j^2-1),
\end{align}
which is of Riccati type. 
Thus, the off-circle branches are genuine solutions of the complexified $X_j$ dynamics, although they do not correspond to real phase differences. Related complex Riccati-type dynamics has recently been studied in Refs.~\cite{cestnik2024integrability,pazo2025spiking}.

\subsection{Phase-amplitude reduction and an invariant}
\label{sec:reduction}

We now focus on how the Koopman eigenfunctions obtained above characterize the dynamics in terms of phase, amplitude, and invariant components~\cite{Wilson2016isostable,shirasaka2017phase}.

Since the mean phase evolves with constant angular velocity as $\dot{\Theta}=\omega$, the phase eigenfunction $\psi_{\mathrm{ph}}=e^{i\Theta}$ represents the one-dimensional phase dynamics.
For the nontrivial part, the eigenfunctions $\psi_1$, $\psi_2$, and $\psi_3$ have the common eigenvalue $-K$ and satisfy $\psi_1+\psi_2+\psi_3=0$.
As noted in Sec.~\ref{sec:eigenfunctions_3KM}, the ratio $\psi_{\mathrm{inv}}={\psi_1}/{\psi_2}$ is a nontrivial invariant. In the $(\psi_1,\psi_2)$ representation, only the overall scale changes exponentially, while the ratio is preserved.

To introduce a nonnegative amplitude-like coordinate preserving the symmetry among $\psi_{1,2,3}$, we define
\begin{align}
    \psi_{\mathrm{amp}} = \sqrt[3]{|\psi_1\psi_2\psi_3|}.
\end{align}
For $K>0$, it satisfies
\begin{align}
    \frac{d\psi_{\mathrm{amp}}}{dt} = -K\psi_{\mathrm{amp}}\leq0,
\end{align}
and can therefore also be regarded as a Lyapunov-like function~\cite{Mauroy2016global}.
Hence, the nontrivial dynamics can be decomposed into an amplitude-like coordinate and an invariant.

Accordingly, the dynamics can be characterized by the phase eigenfunction $\psi_{\mathrm{ph}}$, the amplitude variable $\psi_{\mathrm{amp}}$, and the invariant $\psi_{\mathrm{inv}}$.
In these variables, the dynamics takes the linear form
\begin{align}
\label{eq:linear_ode_reduced}
    \frac{d}{dt}
\begin{pmatrix}
 \psi_{\mathrm{ph}} \\ \psi_{\mathrm{amp}} \\ \psi_{\mathrm{inv}}
\end{pmatrix}
 =
 \begin{pmatrix}
 i\omega & 0 & 0\\
 0 & -K & 0\\
 0 & 0 & 0
\end{pmatrix}
\begin{pmatrix}
 \psi_{\mathrm{ph}} \\ \psi_{\mathrm{amp}} \\ \psi_{\mathrm{inv}}
\end{pmatrix}.
\end{align}

{We note that the Koopman eigenvalue $i\omega$ corresponds to the Floquet exponent~\cite{shirasaka2017phase} associated with the phase direction, whose Floquet multiplier is the neutral multiplier $1$.
Similarly, the common Koopman eigenvalue $-K$ of the nontrivial eigenfunctions coincides with the two transverse Floquet exponents of the synchronized periodic orbit.}

\begin{figure}[htbp]
    \centering
    \includegraphics[width=0.45\linewidth]{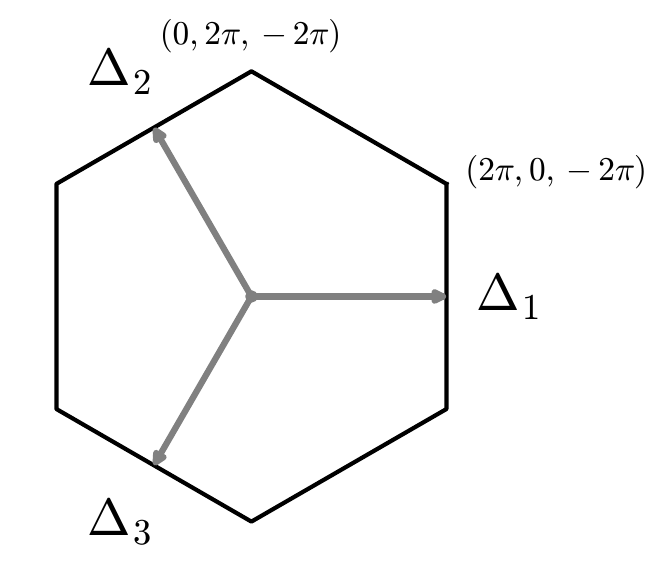}
    \includegraphics[width=0.45\linewidth]{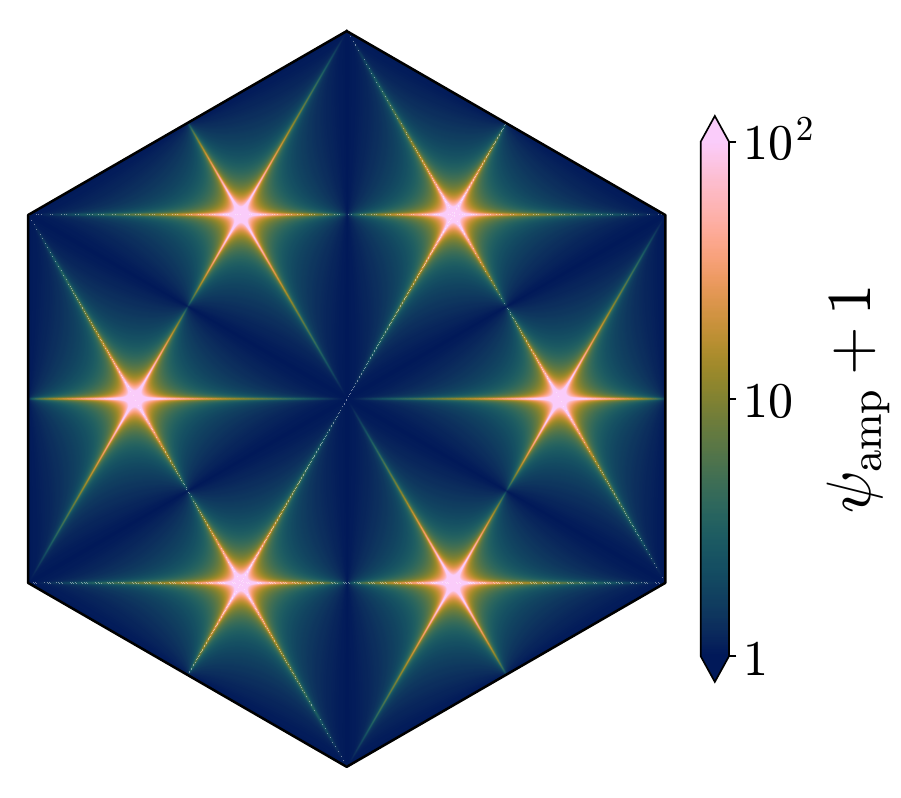}
    \\
    (a-1)\hspace{6cm}(a-2)\\
    \includegraphics[width=0.45\linewidth]{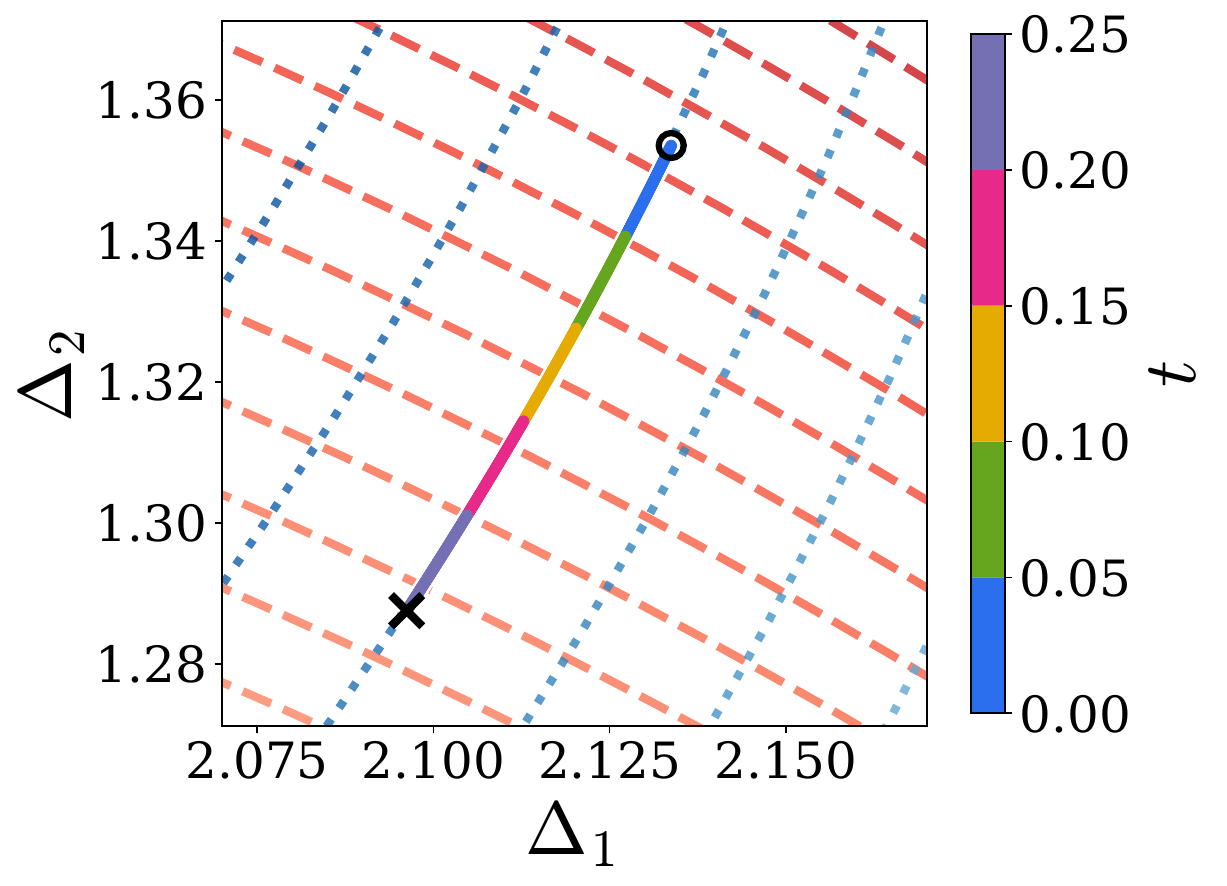}
    \includegraphics[width=0.45\linewidth]{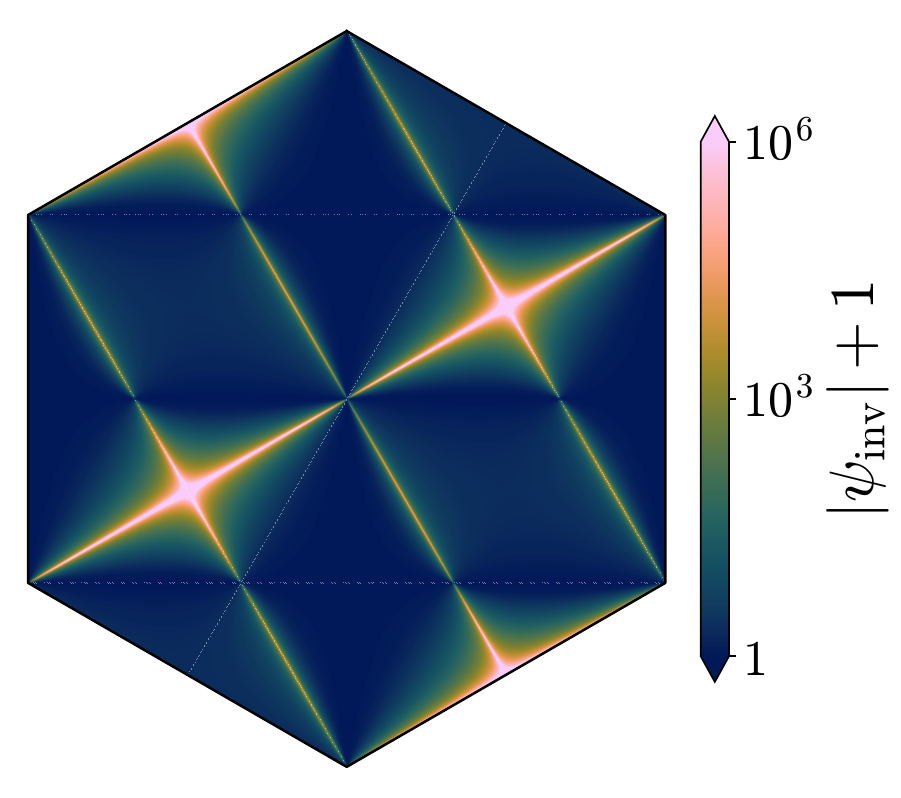}
    \\
    (b)\hspace{6cm}(a-3)
\caption{(a) Variables associated with the Koopman-eigenfunction representation. We plot $\psi_{\mathrm{amp}}+1$ and $|\psi_{\mathrm{inv}}|+1$ on logarithmic color scales to avoid zero values.
(a-1) The projected phase-difference domain. Since $\Delta_1+\Delta_2+\Delta_3=0$, quantities can be represented on a two-dimensional hexagonal region. The outer hexagon is the phase-difference domain induced by $\theta_j\in[0,2\pi]$.
(a-2) The amplitude variable $\psi_{\mathrm{amp}}$ with values clipped to $[0,10^2]$.
The peaks indicate singularities at the splay states.
(a-3) The invariant $\psi_{\mathrm{inv}}$. Here we plot the absolute value $|\psi_{\mathrm{inv}}|$ to show its scale, with values clipped to $[0,10^6]$.
(b) Enlarged view of the contour lines of $\psi_{\mathrm{amp}}$ (red dashed lines) and $\psi_{\mathrm{inv}}$ (blue dotted lines), both drawn at exponentially spaced levels, together with an example trajectory of the identical Kuramoto model for $N=3$ with $K=1$.
The cross marks the starting point, and the circle marks the end point. The trajectory remains on a single contour of the invariant and crosses the contour lines of the amplitude function at regular time intervals.}
    \label{fig:coordinate}
\end{figure}

Figure~\ref{fig:coordinate}(a-1) shows the projected phase-difference domain.
Since $\Delta_1+\Delta_2+\Delta_3=0$, the phase differences can be represented on a two-dimensional hexagonal region.
Figure~\ref{fig:coordinate}(a-2) shows the amplitude function $\psi_{\mathrm{amp}}$ on this domain.
The six peaks correspond to singularities at the splay states, where the denominators of all three Koopman eigenfunctions vanish.
Here, the splay states are phase-locked states with phases separated by $2\pi/3$.
For $K>0$, trajectories starting inside the hexagon formed by the six peaks approach the synchronized state $(0,0,0)$, whereas those starting outside it approach states equivalent to synchrony modulo $2\pi$, such as $(2\pi,-2\pi,0)$.
For $K<0$, the stability of the synchronized and splay states is reversed, and trajectories in the triangular regions around the peaks approach the corresponding splay states.

Figure~\ref{fig:coordinate}(a-3) shows the absolute value of the invariant $\psi_{\mathrm{inv}}=\psi_1/\psi_2$ to visualize its scale.
The plots for $\psi_2/\psi_3$ and $\psi_3/\psi_1$ are obtained from this plot by rotations of $2\pi/3$ and $4\pi/3$, respectively.

Figure~\ref{fig:coordinate}(b) shows the contour lines of $\psi_{\mathrm{amp}}$ and $\psi_{\mathrm{inv}}$, both drawn at exponentially spaced levels, together with an example trajectory.
The trajectory is constrained to a single level set of the invariant and crosses the exponentially spaced contour lines of the amplitude function at regular time intervals.

\section{Extensions}
\label{sec:extensions}
{
\subsection{Extension to the Kuramoto--Sakaguchi model}
\label{sec:kuramoto-sakaguchi}
The Kuramoto--Sakaguchi model extends the standard Kuramoto model by introducing a phase-lag parameter $\alpha$ and is given by
\begin{align}
\dot{\theta}_j = \omega + \frac{K}{N}\sum_{k=1}^{N}
\sin(\theta_k-\theta_j+\alpha).
\end{align}
Here, $\alpha=0$ corresponds to the standard Kuramoto model, whereas $\alpha=\pm\pi/2$ gives pure cosine coupling.

We note that, for the all-to-all coupled Kuramoto--Sakaguchi model, the phase eigenfunction $\psi_{\mathrm{ph}}=e^{i\Theta}$ is recovered at $\alpha = 0$. When $\sin\alpha\neq0$, the mean-phase growth rate becomes state dependent, so this phase exponential is no longer a Koopman eigenfunction. This is also consistent with the general conditions for monomial Koopman eigenfunctions discussed in Ref.~\cite{thibeault2025kuramoto}.

For arbitrary $N$, the pure cosine-coupling case admits a first integral~\cite{Watanabe1993integrability,Watanabe1994constants},
\begin{align}
I_{\mathrm S}^{(N)}
=
\prod_{j=1}^{N}
\sin\frac{\theta_j-\theta_{j+1}}{2},
\qquad
\theta_{N+1}=\theta_1.
\end{align}
The growth-rate approach provides a simple connection for general $\alpha$. We introduce the Kuramoto order parameter
\begin{align}
r=\left|\frac{1}{N}\sum_{j=1}^{N}e^{i\theta_j}\right|.
\end{align}
The growth rates of $I_{\mathrm S}^{(N)}$ and
$\psi_{\mathrm{ph}}=e^{i\Theta}$ are given by
\begin{align}
\frac{\mathcal{K}I_{\mathrm S}^{(N)}}{I_{\mathrm S}^{(N)}}
=
-NKr^2\cos\alpha,\qquad
\frac{\mathcal{K}\psi_{\mathrm{ph}}}{\psi_{\mathrm{ph}}}
=
i\omega+iKr^2\sin\alpha.
\end{align}
Therefore,
\begin{align}
\psi_{\alpha}^{(N)}
=
(\psi_{\mathrm{ph}})^{N\cos\alpha}
\left(I_{\mathrm S}^{(N)}\right)^{i\sin\alpha}
\end{align}
is a Koopman eigenfunction satisfying
\begin{align}
\mathcal{K}\psi_{\alpha}^{(N)}
=
iN\omega\cos\alpha\,
\psi_{\alpha}^{(N)}.
\end{align}
{Here, the complex powers are defined by $(\psi_{\mathrm{ph}})^{N\cos\alpha}=e^{iN\Theta\cos\alpha}$ and $(I_{\mathrm S}^{(N)})^{i\sin\alpha}=e^{i\sin\alpha\log I_{\mathrm S}^{(N)}}$, where the logarithmic branch is chosen continuously on each connected region with $I_{\mathrm S}^{(N)}\neq0$. For general $\alpha$, $\psi_{\alpha}^{(N)}$ is not defined on the set $I_{\mathrm S}^{(N)}=0$, but this set is invariant under the dynamics.
Moreover, a shift $\theta_j\to\theta_j+2\pi$ multiplies $(\psi_{\mathrm{ph}})^{N\cos\alpha}$ by $e^{2\pi i\cos\alpha}$ while leaving $I_{\mathrm S}^{(N)}$ unchanged; hence, $\psi_{\alpha}^{(N)}$ must be regarded as a function of the unwrapped phases $\theta_j\in\mathbb{R}$, and it can be treated modulo $2\pi$ only for $\alpha=0,\pi,\pm\pi/2$.}

In the rotating frame defined by $\theta_j\to\theta_j-\omega t$, this Koopman eigenfunction is equivalent to the constant of motion obtained in Ref.~\cite{ha2021frustration}.

At $\alpha=0$, this reduces to $(\psi_{\mathrm{ph}})^N=e^{iN\Theta}$.
At $\alpha=\pm\pi/2$, its eigenvalue vanishes and $\psi_{\alpha}^{(N)}=(I_{\mathrm S}^{(N)})^{\pm i}$, where $I_{\mathrm S}^{(N)}$ is the first integral of the pure cosine-coupling case.

Although this construction yields a Koopman eigenfunction for arbitrary $N$ and general $\alpha$, it does not directly provide the additional eigenfunctions required for the trajectory reconstruction developed in this paper.

To construct additional Koopman eigenfunctions, we now return to the $N=3$ case. A key step in the present analysis for the $N=3$ standard Kuramoto model is that the phase-difference dynamics itself has the structure of an identical-oscillator system, namely, it takes the same form for each $j$ as in Eq.~\eqref{eq:diff_KM}. This allows a second application of Lemma~\ref{lem:pointwise} and yields simple growth rates. For general $\alpha$, however, the phase-difference dynamics becomes
\begin{align}
\dot{\Delta}_j =\cos\alpha \left[G(\bm{\Delta})-K\sin\Delta_j\right] + \frac{K}{3}\sin\alpha
\left(\cos\Delta_{j+2}-\cos\Delta_{j+1}\right),
\end{align}
where the indices are understood cyclically. In general, the pairwise difference dynamics of the $\Delta_j$ does not satisfy Eq.~\eqref{eq:identical_diff}. Thus, the construction of additional Koopman eigenfunctions for general $\alpha$ requires finding other observables with simple growth rates and remains an open problem.}

\subsection{Extensions to larger systems}
\label{sec:beyond_N3}

Another natural extension is to consider systems with $N\geq4$.
{
The $N=3$ system considered here is special in that the all-to-all coupled network is also the smallest ring network.
This suggests two natural directions for extending the present analysis to larger systems, namely, all-to-all coupling and ring coupling.}

\paragraph{All-to-all coupling}

For $N=4$, the phase-difference dynamics no longer has the identical-oscillator form.
Nevertheless, suitable pairs of phase-difference variables satisfy Eq.~\eqref{eq:identical_diff}, so that Lemma~\ref{lem:pointwise} remains applicable.
We introduce the phase differences
\begin{align}
\Delta_1=\theta_1-\theta_2,\quad
\Delta_2=\theta_4-\theta_3,\quad
\Delta_3=\theta_1-\theta_3,\cr
\Delta_4=\theta_2-\theta_4,\quad
\Delta_5=\theta_1-\theta_4,\quad
\Delta_6=\theta_3-\theta_2.
\end{align}
Applying Lemma~\ref{lem:pointwise} to suitable pairs, we obtain a
nontrivial Koopman eigenfunction with eigenvalue $-K$,
\begin{align}
    \psi_{\mathrm{amp}}^{(4,\mathrm{all})}
    =
    \frac{
    \sin^2\frac{\Delta_1}{2}
    \sin^2\frac{\Delta_2}{2}
    }{
    \sin\frac{\Delta_1-\Delta_2}{2}
    \sin\frac{\Delta_3-\Delta_4}{2}
    \sin\frac{\Delta_5-\Delta_6}{2}
    },
\end{align}
which satisfies $\mathcal K\psi_{\mathrm{amp}}^{(4,\mathrm{all})}=-K\psi_{\mathrm{amp}}^{(4,\mathrm{all})}$.
Together with the mean-phase relation and the cross ratio
\begin{align}
\psi_{\mathrm{cr}} = \frac{
\sin\frac{\Delta_1}{2}\,
\sin\frac{\Delta_2}{2}
}{
\sin\frac{\Delta_5}{2}\,
\sin\frac{\Delta_6}{2}
},
\end{align}
which is related to the Watanabe--Strogatz invariant
~\cite{Marvel2009identical},
this gives three relations among the four phase variables.

{In addition, we obtain a further nontrivial invariant,
\begin{align}
    \psi_{\mathrm{inv}}^{(4,\mathrm{all})} = \left({\frac{1-x}{1+x}}\right)^{\frac{\psi_{\mathrm{cr}}+1}{\psi_{\mathrm{cr}}}} \left({\frac{1+y}{1-y}}\right)^{\frac{1}{\psi_{\mathrm{cr}}}}\frac{(\psi_{\mathrm{cr}}+1)x-y+\psi_{\mathrm{cr}}}{(\psi_{\mathrm{cr}}+1)x-y-\psi_{\mathrm{cr}}},
\end{align}
where
\begin{align}
x=\cos\frac{\Delta_1-\Delta_2}{2},\qquad
y=\cos\frac{\Delta_3-\Delta_4}{2}.
\end{align}
A direct differentiation along the flow verifies $\mathcal K\psi_{\mathrm{inv}}^{(4,\mathrm{all})}=0$.
Together with the phase mode, these quantities provide four Koopman-based relations for the $N=4$ system.
Unlike the $N=3$ case, however, their inversion is generically transcendental because the exponents in
$\psi_{\mathrm{inv}}^{(4,\mathrm{all})}$ depend on the invariant $\psi_{\mathrm{cr}}$ and are generally noninteger.
Thus, the algebraic reconstruction available for $N=3$ does not directly extend to $N=4$.}

For $N\geq5$, one phase eigenfunction and the $N-3$ cross-ratio invariants remain available, but we have not found two additional Koopman eigenfunctions that provide the missing relations required for explicit trajectory reconstruction.

{
\paragraph{Ring coupling}

For the $N=4$ nearest-neighbor ring, the model is given by
\begin{align}
    \dot{\theta}_j = \omega + \frac{K}{4}\left(\sin(\theta_{j-1} - \theta_j) + \sin(\theta_{j+1} - \theta_j)\right),\qquad j=1,2,3,4.
\end{align}
Because the coupling is symmetric, the interaction terms cancel in the evolution of the mean phase $\Theta$. Thus, $\dot{\Theta}=\omega$, and $\psi_{\mathrm{ph}}=e^{i\Theta}$ is a Koopman eigenfunction with eigenvalue $i\omega$.

Beyond this phase mode, the growth-rate construction can also be applied to the phase dynamics.
Within each pair $(\theta_1,\theta_3)$ and $(\theta_2,\theta_4)$, the oscillators obey equations of the same form, so that their pairwise difference dynamics satisfies Eq.~\eqref{eq:identical_diff}.
This gives
\begin{align}
\psi_{\mathrm{inv},1}^{(4,\mathrm{ring})}
=
\frac{
\sin\frac{\Delta_3}{2}
}{
\sin\frac{\Delta_4}{2}
},
\end{align}
which is a previously known first integral
~\cite{Burylko2022Symmetry,thibeault2025kuramoto}.

Furthermore, because the relevant phase-difference variables $\Delta_1$ and $\Delta_2$ obey equations of the same form, the present growth-rate argument yields a second Koopman eigenfunction with zero eigenvalue,
\begin{align}
\psi_{\mathrm{inv},2}^{(4,\mathrm{ring})}
&=
\frac{
\sin\frac{\Delta_2-\Delta_1}{2}
}{
\sin\frac{\Delta_3}{2}
\sin\frac{\Delta_4}{2}
}.
\end{align}

Let $\eta=\Delta_2-\Delta_1$ and use $(\Delta_3,\Delta_4,\eta)$ as independent coordinates. A direct calculation gives
\begin{align}
\det
\frac{
\partial
\left(
\psi_{\mathrm{inv},1}^{(4,\mathrm{ring})},
\psi_{\mathrm{inv},2}^{(4,\mathrm{ring})}
\right)
}{
\partial(\Delta_3,\eta)
}
=
\frac{
\cos\frac{\Delta_3}{2}
\cos\frac{\eta}{2}
}{
4
\sin\frac{\Delta_3}{2}
\sin^2\frac{\Delta_4}{2}
}.
\end{align}
This determinant is nonzero on a generic regular region.
Hence, the two first integrals are functionally independent there.
Fixing the values of both invariants reduces the three-dimensional phase-difference dynamics to one-dimensional motion, so the $N=4$ ring is completely integrable by quadrature.

A further Koopman eigenfunction can in principle be constructed from the corresponding time coordinate.
This complete integrability, however, does not by itself provide a simple closed-form reconstruction of the individual phase trajectories.

Comparison with the all-to-all coupled system shows that the Koopman structure relevant to trajectory reconstruction depends not only on system size but also on coupling topology.

For ring systems with $N\geq5$, the phase Koopman eigenfunction remains available, but we have not found analogous nontrivial Koopman eigenfunctions that provide sufficient relations for trajectory reconstruction. Extensions to more general systems with non-global interactions are also of interest~\cite{thibeault2025kuramoto}.
}

%%%%%%%%%%%%%%%%%%%%%%%%%%%%%%%%%%%
\section{Conclusion}
In this study, we considered the identical Kuramoto model with $N=3$, the simplest nontrivial case for which global exact trajectories had not previously been obtained.

We identified Koopman eigenfunctions corresponding to the phase, amplitude, and invariant components of the dynamics. Using them as relations between the state variables and time, we reduced trajectory reconstruction to solving time-dependent quartic equations and selecting the algebraic branch corresponding to the initial condition.
This provides a global analytical solution for the phase trajectories of the identical Kuramoto model with $N=3$.

The Watanabe--Strogatz transformation provides a three-dimensional reduction of the essential dynamics of identical oscillator systems.
For $N=2$, this reduced description is redundant, whereas $N=3$ is the smallest nonredundant realization of the Watanabe--Strogatz reduction.
The present results show that, in this minimal case, the reduced dynamics can be reconstructed explicitly through algebraic relations generated by Koopman eigenfunctions.

{The extensions considered here also illustrate how this structure changes beyond the $N=3$ standard Kuramoto model.
For arbitrary $N$ and general phase lag in the all-to-all coupled Kuramoto--Sakaguchi model, an explicit Koopman eigenfunction can still be constructed, although even for $N=3$ the additional structure used here for trajectory reconstruction is no longer directly available away from pure sine coupling.

For the $N=4$ all-to-all system, four Koopman-based relations are obtained, but their inversion is generically transcendental rather than algebraic.
In contrast, the $N=4$ ring possesses two independent invariants and is completely integrable by quadrature.
These examples indicate that the feasibility of explicit trajectory reconstruction depends both on system size and on coupling topology.
Whether similarly explicit global analytical reconstructions can be obtained for larger oscillator systems remains an open question.

The exact Koopman eigenfunctions and trajectories obtained here may also provide an analytical benchmark for numerical Koopman-based methods~\cite{brunton2022modern}, such as dynamic mode decomposition, in nonlinear interacting oscillator systems.}

\section*{Acknowledgments}
The author is grateful to Hiroya Nakao and Iv\'an Leon for valuable discussions and comments. The author also thanks the anonymous reviewers for their useful comments. This work was supported by JSPS KAKENHI Grant Number 24K20863.

\section*{AUTHOR DECLARATIONS}

\subsection*{Conflict of Interest}
The author has no conflicts to disclose.

\section*{DATA AVAILABILITY}
The data that support the findings of this study are available within the article.

%%%%%%%%%%%%%%%%%%%%%%%%%%%%%%
% \clearpage
\appendix
\section{Proof of Lemma~\ref{lem:pointwise}}
\label{sec:proof_pointwise}
The lemma follows by direct calculation.
\begin{align}
    \Lambda=&\frac{\mathcal{K}u}{u} = \frac{1}{2}[g(\cos\varphi_j-\cos\varphi_k) + h(\sin\varphi_j-\sin\varphi_k)]\cot \frac{\varphi_j - \varphi_k}{2}\cr
    =& -g\sin\frac{\varphi_j+\varphi_k}{2}\cos\frac{\varphi_j-\varphi_k}{2} + h\cos\frac{\varphi_j+\varphi_k}{2}\cos\frac{\varphi_j-\varphi_k}{2}\cr
    =& -\frac{g}{2}(\sin\varphi_j+\sin\varphi_k) + \frac{h}{2}(\cos\varphi_j+\cos\varphi_k).
\end{align}
This proves the claim.

\section{Koopman eigenfunctions for the identical Kuramoto model with $N=3$}
\subsection{Derivation of the Koopman eigenfunctions}
\label{sec:eig_deriv}

Since the identical Kuramoto model~\eqref{eq:3kuramoto} is a special case of the identical oscillator system~\eqref{eq:identical_oscillators}, we first consider the following observables:
\begin{align}
    u_1 = \sin\frac{\theta_1-\theta_2}{2},\quad u_2 = \sin\frac{\theta_2-\theta_3}{2},\quad u_3 = \sin\frac{\theta_3-\theta_1}{2}.
\end{align}
Using Lemma~\ref{lem:pointwise}, we obtain the corresponding growth rates $\Lambda_{1,2,3}$:
\begin{align}
    \Lambda_1 = -\frac{K}{6}\sum_{k=1}^3[\cos(\theta_k-\theta_1) + \cos(\theta_k-\theta_2)],\cr
    \Lambda_2 = -\frac{K}{6}\sum_{k=1}^3[\cos(\theta_k-\theta_2) + \cos(\theta_k-\theta_3)],\cr
    \Lambda_3 = -\frac{K}{6}\sum_{k=1}^3[\cos(\theta_k-\theta_3) + \cos(\theta_k-\theta_1)].\cr
    \label{eq:1st_n=3}
\end{align}
To simplify the expressions, we introduce the phase differences:
\begin{align}
    \Delta_1=\theta_1-\theta_2,\quad \Delta_2=\theta_2-\theta_3,\quad \Delta_3=\theta_3-\theta_1.
\end{align}
Expressed in terms of the phase differences, Eq.~\eqref{eq:1st_n=3} can be written as
\begin{align}
    \Lambda_j = -\frac{K}{6}(2+\cos\Delta_1+\cos\Delta_2+\cos\Delta_3 + \cos\Delta_j).
\end{align}

Next, the dynamics of the phase differences $\Delta_j$ are given by
\begin{align}
\label{eq:phase_difference3}
    \dot{\Delta}_j = G(\Delta_1,\Delta_2,\Delta_3) - K \sin\Delta_j,
\end{align}
where $G(\Delta_1,\Delta_2,\Delta_3) = \frac{K}{3}(\sin\Delta_1 + \sin\Delta_2+\sin\Delta_3)$ is common to all $\Delta_j$.
Thus, Eq.~\eqref{eq:phase_difference3} also belongs to the class of the identical oscillator system~\eqref{eq:identical_oscillators}, so Lemma~\ref{lem:pointwise} applies.
Therefore, we further consider the following observables:
\begin{align}
    u_4 =& \sin\left(\frac{\Delta_1-\Delta_2}{2}\right),\quad u_5 = \sin\left(\frac{\Delta_2-\Delta_3}{2}\right),\cr
    u_6 =& \sin\left(\frac{\Delta_3-\Delta_1}{2}\right).
\end{align}
Applying the lemma again, we obtain the corresponding growth rates $\Lambda_{4,5,6}$:
\begin{align}
    \Lambda_4=& -\frac{K}{2}(\cos\Delta_1 + \cos\Delta_2), \quad \Lambda_5= -\frac{K}{2}(\cos\Delta_2 + \cos\Delta_3), \cr
    \Lambda_6=& -\frac{K}{2}(\cos\Delta_3 + \cos\Delta_1).
\end{align}

Combining the above results and choosing the coefficients as follows:
\begin{align}
    \bm{\alpha}_1 =& (3,0,0,-1,0,-1),\quad \bm{\alpha}_2 = (0,3,0,-1,-1,0),\cr
    \bm{\alpha}_3 =& (0,0,3,0,-1,-1),
\end{align}
it follows that $\sum_{k=1}^6 \Lambda_k \alpha_{j,k} = -K$ for $j=1,2,3$, where $\alpha_{j,k}$ denotes the $k$th component of $\bm{\alpha}_j$.
Thus, we obtain Koopman eigenfunctions $\psi_{1,2,3}$ in Eq.~\eqref{eq:koopman_eigenfunction_phase_original} with eigenvalue $-K$.

%%%%%%%%%%%%%%%%%%%%%%%%%%
{
\subsection{Proof of $\psi_1+\psi_2+\psi_3=0$}
\label{sec:proof_sum}
The three Koopman eigenfunctions obtained above satisfy the linear relation $\psi_1+\psi_2+\psi_3=0$ and are therefore not independent.
We write
\begin{align}
F_d=
\sin\frac{\Delta_1-\Delta_2}{2}
\sin\frac{\Delta_2-\Delta_3}{2}
\sin\frac{\Delta_3-\Delta_1}{2}.
\end{align}
Then
\begin{align}
\psi_1+\psi_2+\psi_3
=
\frac{F_n}{F_d},
\end{align}
where
\begin{align}
F_n=
\sum_{j=1}^3
\sin^3\frac{\Delta_j}{2}
\sin\frac{\Delta_{j+1}-\Delta_{j+2}}{2}.
\end{align}
Thus, it is sufficient to show that $F_n=0$.

Using the trigonometric formula,
\begin{align}
4\sin^3\frac{\Delta_j}{2} = 3\sin\frac{\Delta_j}{2} - \sin\frac{3\Delta_j}{2}
\end{align}
and $\Delta_1+\Delta_2+\Delta_3=0$, we obtain
\begin{align}
8F_n = \sum_{j=1}^3
\Bigl[
3\bigl(\cos\Delta_{j+1}-\cos\Delta_{j+2}\bigr) - \cos(\Delta_j-\Delta_{j+1}) + \cos(\Delta_j-\Delta_{j+2})
\Bigr].
\end{align}
Both terms vanish under the cyclic sum, and hence $F_n=0$. Therefore,
\begin{align}
\psi_1+\psi_2+\psi_3=0.
\end{align}

\subsection{Local invertibility of the Koopman coordinates}
\label{sec:invertibility}

We show that the Koopman coordinates used in the reconstruction are
functionally independent on a generic regular region.
Since $\psi_{\mathrm{ph}}=e^{i\Theta}$ takes values on the unit circle,
we use $\Theta$ as a local real coordinate for $\psi_{\mathrm{ph}}$.
Thus, it is sufficient to examine the map
\begin{align}
(\theta_1,\theta_2,\theta_3)
\mapsto
(\Theta,\psi_1,\psi_2).
\end{align}

Using
\begin{align}
\Theta=\frac{\theta_1+\theta_2+\theta_3}{3},
\qquad
\Delta_1=\theta_1-\theta_2,
\qquad
\Delta_2=\theta_2-\theta_3,
\end{align}
with $\Delta_3=-\Delta_1-\Delta_2$, and noting that
\begin{align}
\det
\frac{\partial(\Theta,\Delta_1,\Delta_2)}
{\partial(\theta_1,\theta_2,\theta_3)}
=1,
\end{align}
a direct calculation gives
\begin{align}
\det \frac{\partial(\Theta,\psi_1,\psi_2)}
{\partial(\theta_1,\theta_2,\theta_3)} = -\frac{3}{2}
\frac{\prod_{j=1}^3 \sin^2(\Delta_j/2)}{\prod_{j<k}
\sin^2\left((\Delta_j-\Delta_k)/2\right)}.
\end{align}
Hence, away from the sets $\sin(\Delta_j/2)=0$ and
$\sin((\Delta_j-\Delta_k)/2)=0$, the Jacobian determinant is finite and nonzero.
The inverse function theorem therefore implies that
$(\Theta,\psi_1,\psi_2)$ provides locally invertible coordinates on a
generic regular region. Equivalently,
$(\psi_{\mathrm{ph}},\psi_1,\psi_2)$ is locally invertible on each
regular phase branch.

Since these coordinates satisfy $\dot{\Theta}=\omega$ and $\dot{\psi}_{1,2}=-K\psi_{1,2}$ along the flow of the Kuramoto model, their local inverse reconstructs the same local trajectory of Eq.~\eqref{eq:3kuramoto}.
The exceptional cases are treated separately in Appendices~\ref{sec:degenerate_cases} and~\ref{sec:singular_cases}.
}
%%%%%%%%%%%%%%%%%%%%%%%%%%
\section{Quartic reconstruction with finite Koopman eigenfunctions}
\label{sec:quartic_roots}
{In this appendix, we consider the region where the Koopman eigenfunctions remain finite. This includes degenerate cases in which the Koopman-coordinate map is not locally invertible or the quartic equations have multiple roots.
Throughout Appendices~\ref{sec:quartic_roots} and~\ref{sec:singular_cases}, since the dynamics is invariant under independent shifts $\theta_j\mapsto\theta_j+2\pi n_j$ ($n_j\in\mathbb Z$), we choose without loss of generality the initial phases $\theta_j(0)$ modulo $2\pi$ so that the equalities among the phase differences in each case hold exactly, and compute $\Theta$ in Eq.~\eqref{eq:phase_rel} from these representatives.}

\subsection{Unit-circle roots and the discriminant}

We examine the roots of the quartic polynomials in Eq.~\eqref{eq:quartic_R} on the unit circle.
For this purpose, we apply the Cayley transformation given by
\begin{align}
\label{eq:cayley}
    c_j = \cot\frac{\Delta_j}{2} = i\frac{X_j+1}{X_j-1}
    \quad\leftrightarrow\quad
    X_j = \frac{c_j+i}{c_j-i}.
\end{align}
This transformation maps the unit circle, except for $X_j=1$, to the real
axis. Applying it to $R_j(X_j)$, we obtain a new polynomial $\widetilde{R}_j(c_j)$ as
\begin{align}
    R_j(X_j) = \frac{16\widetilde{R}_j(c_j)}{(c_j-i)^4}.
\end{align}
If $X_j$ lies on the unit circle and $X_j\neq 1$, then $c_j$ is real-valued; in particular, $c_j=i$ does not occur.

The transformed polynomial $\widetilde{R}_j(c_j)$ can be written as
\begin{align}
\label{eq:quartic_cayley}
\widetilde R_j(c_j) = b_{j,4}c_j^4 + b_{j,3}c_j^3 + b_{j,2}c_j^2 + b_{j,1}c_j + b_{j,0}.
\end{align}
Using the constraint $\psi_j+\psi_{j+1}+\psi_{j+2}=0$, the coefficients in Eq.~\eqref{eq:quartic_cayley} are given by
\begin{align}
b_{j,4} =& -9\psi_j^2\psi_{j+1}\psi_{j+2},\cr
b_{j,3} =& -4\psi_j(2\psi_{j+1}+\psi_{j+2})(\psi_{j+1}+2\psi_{j+2}),\cr
b_{j,2} =& 6\psi_j^2\left(\psi_{j+1}\psi_{j+2}-2\right),\cr
b_{j,1} =& 6\psi_j\left(2\psi_{j+1}\psi_{j+2}-1\right),\cr
b_{j,0} =& -\left(\psi_j^2\psi_{j+1}\psi_{j+2}-4\psi_{j+1}\psi_{j+2}+1\right).
\end{align}

For real phase differences, the Koopman eigenfunctions $\psi_j$ are real-valued.
Hence, $\widetilde R_j(c_j)$ has real-valued coefficients.
Since the Cayley transformation maps the unit circle except for $X_j=1$ to the real axis, real roots of $\widetilde R_j(c_j)$ are in one-to-one correspondence with unit-circle roots of $R_j(X_j)$ other than $X_j=1$.
The number of real roots in the generic case can be inferred from the discriminant.
The discriminant of $\widetilde R_j(c_j)$, $D_{\widetilde R_j}$, factorizes as
\begin{align}
\label{eq:discriminant}
D_{\widetilde R_j} = -432\psi_j^4\psi_{j+1}^2\psi_{j+2}^2(\psi_{j+1}-\psi_{j+2})^4(4\psi_j^2+1)^3(4\psi_{j+1}^2+1)(4\psi_{j+2}^2+1).
\end{align}

Hence, the discriminant is non-positive, and it is negative in the generic case.
Since a real quartic polynomial with negative discriminant has exactly two real roots and one pair of non-real complex conjugate roots, $\widetilde R_j(c_j)$ has two distinct real roots in the generic case. Multiple roots occur only in the exceptional cases where $D_{\widetilde R_j}=0$. Via the Cayley transformation, these real roots correspond to roots of $R_j(X_j)$ on the unit circle.

Within the regular formulation, the degenerate cases are classified into two types. The first type corresponds to vanishing Koopman eigenfunctions, such as $\psi_1=0$, $\psi_2=0$, and $\psi_3=0$.
The second type occurs when $\psi_{j+1}=\psi_{j+2}$, as indicated by the factor $(\psi_{j+1}-\psi_{j+2})^4$ in Eq.~\eqref{eq:discriminant}.
For example, $R_1(X_1)$ becomes degenerate when $\psi_2=\psi_3$.

The discriminant analysis above concerns the regular part, where the relevant Koopman eigenfunctions remain finite and the quartic formulation is valid.
Singular cases, where some denominators of the Koopman eigenfunctions vanish, are discussed in Appendix~\ref{sec:singular_cases} (summarized in Table~\ref{tab:cases}).

\subsection{Degenerate cases}
\label{sec:degenerate_cases}
We now analyze these two degenerate but regular cases explicitly.
The quartic formulation remains valid because the Koopman
eigenfunctions are finite.

\subsubsection*{Case 1: $\psi_j=0$ and $\psi_{j+1}=-\psi_{j+2}\neq0$}

The relation $\psi_{j+1}=-\psi_{j+2}$ follows from $\psi_j+\psi_{j+1}+\psi_{j+2}=0$. By symmetry, it is sufficient to consider the case $\psi_1=0$ and $\psi_2=-\psi_3\neq0$.

Within the regular formulation, $\psi_1=0$ corresponds to $\Delta_1=0 \pmod{2\pi}$, and hence $X_1=1$.
This point is excluded from the cotangent parametrization for $c_1$.
Indeed, substituting $\psi_1=0$ and $\psi_3=-\psi_2$ gives
\begin{align}
\widetilde{R}_1(c_1)=-(4\psi_2^2+1),
\end{align}
which has no root for real $\psi_2$.
Thus, $\widetilde R_1$ does not provide a relation for $c_1$ in this case.
The missing solution is $X_1 = 1$, which belongs to the factor $(X_1-1)^5$ originally contained in $\mathrm{Res}_{X_2}(\mathrm{P1},\mathrm{P2})$.

Substituting $\psi_1=0$ and $\psi_2=-\psi_3\neq0$ into $\widetilde R_2(c_2)$, we obtain a polynomial equation
\begin{align}
    \widetilde{R}_2(c_2) = -8\psi_2^3 c_2^3 -12 \psi_2^2 c_2^2 -6\psi_2c_2 - 1 = -\left(2\psi_2 c_2+1\right)^3 = 0.
\end{align}
Thus, from $2\psi_2 c_2+1=0$, we can recover $\Delta_2$ as
\begin{align}
    c_2=\cot\frac{\Delta_2}{2} = -\frac{1}{2\psi_2}
    \quad\Leftrightarrow\quad \Delta_2 =  -2\arctan(2\psi_2), 
\end{align}
where the branch is chosen consistently with the initial condition.
Therefore, we recover the full dynamics:
\begin{align}
    \theta_1 = \Theta - \frac{2}{3}\arctan(2\psi_2),\quad
    \theta_2 = \Theta - \frac{2}{3}\arctan(2\psi_2),\quad
    \theta_3 = \Theta + \frac{4}{3}\arctan(2\psi_2).
\end{align}

\subsubsection*{Case 2: $\psi_{j+1}=\psi_{j+2}\neq0$}
By symmetry, we consider the degeneracy of $\widetilde R_1(c_1)$, for which $\psi_2=\psi_3$.
We assume $\psi_2=\psi_3\neq0$; the case $\psi_2=\psi_3=0$ is singular and is treated in Appendix~\ref{sec:singular_cases}.
The relation $\psi_1=-2\psi_2$ then follows from $\psi_1+\psi_2+\psi_3=0$.
Substituting these relations into $\widetilde{R}_1(c_1)$, we obtain
\begin{align}
    \widetilde{R}_1(c_1) = -\left( 6\psi_2^2c_1^2 - 6\psi_2c_1 - 2\psi_2^2 +1\right)^2.
\end{align}
Therefore, the remaining condition is
\begin{align}
    6\psi_2^2c_1^2 - 6\psi_2c_1 - 2\psi_2^2 + 1 = 0.
\end{align}
The roots are given by
\begin{align}
    c_1=\frac{3\pm\sqrt{3+12\psi_2^2}}{6\psi_2}.
\end{align}
Then, with the branch chosen consistently with the initial condition, we obtain
\begin{align}
    \Delta_1=2\operatorname{arccot}c_1,
\end{align}
where the branch of $\operatorname{arccot}$ is chosen consistently with the initial condition.
For the remaining variables $c_2$ and $c_3$, the condition $\psi_2=\psi_3$ does not make the corresponding quartic polynomials degenerate. Therefore, their roots are obtained from the corresponding nondegenerate quartic equations.

\section{Singular cases}
\label{sec:singular_cases}

We now consider cases in which at least one denominator of the Koopman eigenfunctions $\psi_{1,2,3}$ vanishes.
The denominators of $\psi_{1,2,3}$ vanish when
\begin{align}
\sin\frac{\Delta_j-\Delta_k}{2}=0
\end{align}
for some pair $(j,k)$. 
Thus, up to permutation symmetry, we may assume
\begin{align}
\Delta_1=\Delta_2 \quad \pmod{2\pi}.
\end{align}
Together with $\Delta_1+\Delta_2+\Delta_3=0 \pmod{2\pi}$, this gives
\begin{align}
(\Delta_1,\Delta_2,\Delta_3)=(\delta,\delta,-2\delta).
\end{align}

In this case, the regular quartic equations $R_j(X_j)$ are no longer directly applicable.
The dynamics needs to be reconstructed separately, as shown below.

\subsection*{Case 1: Fixed points}
Substituting $(\Delta_1,\Delta_2,\Delta_3)=(\delta,\delta,-2\delta)\pmod{2\pi}$ into the phase-difference dynamics~\eqref{eq:diff_KM}, the fixed-point condition reduces to
\begin{align}
\sin \delta=-\sin 2\delta
\quad\leftrightarrow\quad
\sin \delta\ (1+2\cos \delta)=0.
\end{align}
Hence $\delta=0,\pi,\pm 2\pi/3$ modulo $2\pi$, giving the phase-locked states: the synchronized state, the antipodal two-cluster states, and the splay states.

\subsection*{Case 2: $\Delta_j =\Delta_{j+1}\neq\Delta_{j+2}\pmod{2\pi}$ excluding the phase-locked states of Case 1}

By symmetry, it is sufficient to consider $\Delta_1=\Delta_2\neq\Delta_3$.
In this case, the only nontrivial Koopman eigenfunction that remains is $\psi_3$. By substituting $\Delta_1=\Delta_2$ and $\Delta_3=-\Delta_1-\Delta_2=-2\Delta_1$ into $\psi_3$ in Eq.~\eqref{eq:koopman_eigenfunction_phase_original}, we obtain
\begin{align}
\psi_3 = - \frac{\sin^3\frac{\Delta_3}{2}}{\sin^2\frac{\Delta_3-\Delta_1}{2}} = \frac{\sin^3\Delta_1}{\sin^2\frac{3\Delta_1}{2}} = \frac{(X_1^2-1)^3}{2i(X_1^3-1)^2}.
\end{align}

Clearing denominators yields an equation of the form
\begin{align}
(X_1-1)^2 S(\psi_3;X_1) = 0,
\end{align}
where $S(\psi_3;X_1)$ is the quartic polynomial
\begin{align}
    \label{eq:koopman_poly}
    S(\psi_3;X_1) = (2i\psi_3-1)X_1^4+(4i\psi_3-2)X_1^3+6i\psi_3X_1^2+(4i\psi_3+2)X_1+(2i\psi_3+1).
\end{align}
Applying the Cayley transformation~\eqref{eq:cayley} to
$S(\psi_3;X_1)$ gives
\begin{align}
    S(\psi_3;X_1)
    =
    \frac{2i\widetilde S(c_1)}{(c_1-i)^4},
\end{align}
where
\begin{align}
    \widetilde S(c_1)
    =
    9\psi_3c_1^4-8c_1^3-6\psi_3c_1^2+\psi_3.
\end{align}
Thus, roots of $S$ on the unit circle, except for $X_1=1$, correspond to real roots of
\begin{align}
    \widetilde{S}(c_1)
    = 9\psi_3 c_1^4 - 8c_1^3 - 6\psi_3 c_1^2 + \psi_3 = 0.
\end{align}

The discriminant of $\widetilde S$ is
\begin{align}
    \mathrm{D}_{\widetilde S}
    =
    -110592\,\psi_3^2(4\psi_3^2+1).
\end{align}
Hence, for $\psi_3\neq0$, $\widetilde S$ has exactly two distinct real roots and one complex-conjugate pair. The two real roots correspond, through the Cayley transformation, to two unit-circle roots of $S(\psi_3;X_1)$.
The case $\psi_3=0$ belongs to the fixed-point case discussed above.

Once the branch for $X_1$ consistent with the given initial condition has been selected, we recover $\Delta_1=-i\log X_1$ with the appropriate logarithmic branch, and the original phases are given by
\begin{align}
    \theta_1 = \Theta + \Delta_1,\quad \theta_2 = \Theta,\quad \theta_3 = \Theta-\Delta_1.
\end{align}

\section{Quartic equation for the XY Hamiltonian}
\label{sec:xy_quartic}
We derive a quartic equation satisfied by the XY Hamiltonian $H$ in Eq.~\eqref{eq:XY_hamiltonian}.
Introducing the variables $X_j=e^{i\Delta_j}$, where $\Delta_1=\theta_1-\theta_2$, $\Delta_2=\theta_2-\theta_3$, and $\Delta_3=\theta_3-\theta_1$, we obtain the following multi-variable polynomial equation for $H$:
\begin{align}
    2H + JX_1 + JX_2 + JX_3 + JX_1X_2 + JX_3X_1 + JX_2X_3 = 0.
\end{align}
Using the relation $X_1X_2X_3=1$, i.e., $X_3=\frac{1}{X_1X_2}$, we obtain
\begin{align}
    2 H X_1 X_2 + J X_1^2 X_2^2 + J X_1^2 X_2 + J X_1 X_2^2 + J X_1 + J X_2 + J = 0. \tag{Ph1}
\end{align}
We first eliminate $X_2$ using the resultant $\mathrm{Res}_{X_2}(\mathrm{P1},\mathrm{Ph1})$. This yields the following polynomial equation:
\begin{align}
    4 H X_1^2 \psi_{1} + 2 J X_1^4 \psi_{1} - i J X_1^4 + 4 J X_1^3 \psi_{1} + 2 i J X_1^3 + 4 J X_1 \psi_{1} - 2 i J X_1 + 2 J \psi_{1} + i J = 0 \tag{Ph2}
\end{align}
We then consider $\mathrm{Res}_{X_1}(R_1,\mathrm{Ph2})$ and obtain a quartic equation for $H$ of the form
\begin{align}
    P_H(H) = d_4 H^4 + d_3 H^3 + d_2 H^2 + d_1 H + d_0 = 0,
\end{align}
where the coefficients $d_j$ are functions of $\psi_1$ and $\psi_2$. Thus, $H$ is determined algebraically by $\psi_1$ and $\psi_2$. After using $\psi_1+\psi_2+\psi_3=0$, the coefficients can be written as functions of $\psi_1$, $\psi_2$ and $\psi_3$. The coefficients are lengthy; for example, the leading coefficient is
\begin{align}
d_4 = \prod_{j=1}^3(4\psi_j^2+1)^2.
\end{align}

For compactness, we introduce
\begin{align}
    U=\psi_1\psi_2+\psi_2\psi_3+\psi_3\psi_1.
\end{align}
The discriminant is
\begin{align}
\label{eq:disc_PH}
    D_{P_H} =&
-7247757312 J^{12}
\psi_1^4\psi_2^4\psi_3^4
(\psi_1-\psi_2)^4(\psi_2-\psi_3)^4(\psi_3-\psi_1)^4
\prod_{j=1}^3(4\psi_j^2+1)^3\cr
&\qquad\times\left[
6912\psi_1^2\psi_2^2\psi_3^2+(3+4U)^3(4U-1)
\right]^2.
\end{align}

For real phase differences, the Koopman eigenfunctions $\psi_j$ are real-valued, and $P_H(H)$ is a real quartic polynomial.
Since the last factor in Eq.~\eqref{eq:disc_PH} is squared and $4\psi_j^2+1>0$, the discriminant is non-positive.
It is negative in the generic case, where
\begin{align}
   J\neq0,\qquad
   \psi_1\psi_2\psi_3\neq0,\qquad
   \psi_j\neq\psi_k\quad (j\neq k),
\end{align}
and the remaining squared factor in Eq.~\eqref{eq:disc_PH} is nonzero.
Therefore, a negative discriminant implies that $P_H(H)$ has two real roots and one pair of non-real complex conjugate roots.
Thus, the two-sheet structure of the XY energy landscape follows from the two real roots of this quartic equation in the generic case.

\end{document}